\documentclass[12pt]{article}

\usepackage{booktabs}
\PassOptionsToPackage{hypertexnames=false}{hyperref}

\usepackage{jheppub}
\usepackage{amssymb,amsmath,amsthm}
\usepackage[shortlabels]{enumitem}
\usepackage{braket}
\usepackage{comment}
\usepackage{capt-of}
\usepackage{graphicx, xcolor, varwidth}
\usepackage{color,wrapfig}

\newenvironment{figurecaptionblock}
  {\par\noindent\begin{minipage}{0.96\linewidth}\hbadness=2000\relax}
  {\end{minipage}\par}

\DeclareMathOperator{\tr}{tr}

\DeclareMathAlphabet{\mathpzc}{OT1}{pzc}{m}{it}

\renewcommand{\j}{\mathcal{J}}
\renewcommand{\k}{\mathcal{K}}
\newcommand{\h}{\mathbf{H}}

\title{\Large Quantum State of a Gravitating Spacetime Region}
\author{Raphael Bousso, Sami Kaya, Guanda Lin, and Arvin
Shahbazi-Moghaddam}

\affiliation{Leinweber Institute for Theoretical Physics and Department of Physics,
University of California, Berkeley, California 94720, U.S.A.
and \\
Leinweber Institute for Theoretical Physics, Stanford, CA 94305, U.S.A.
}

\emailAdd{bousso@berkeley.edu}
\emailAdd{samikaya@berkeley.edu}
\emailAdd{geoff\_guanda\_lin@berkeley.edu}
\emailAdd{arvinshm@gmail.com}

\abstract{We associate a gravitational Hilbert space $\mathbf{H}_\sigma$ to any closed compact $(d-1)$-manifold $\sigma$ with real metric.
A quantum state $\j(\sigma)$ is a $d$-manifold bounded by $\sigma$ and equipped with elliptic data.
An inner product is defined by gluing states pairwise across $\sigma$ and evaluated by viewing the resulting closed $d$-manifold as a boundary condition on the gravitational path integral (GPI) over $(d+1)$-manifolds.

If $\sigma$ is nonempty and the GPI is dominated by a single $(d+1)$-manifold $M$ in the $G_N\to 0$ limit, then $M$ contains a Lorentzian CRT fixed-point set, providing $\j(\sigma)$ with a classical spacetime interpretation.
Conversely, given a finite Lorentzian domain with edge $\sigma$, a state $\j(\sigma)$ may be associated to it by deforming its initial data off the real Lorentzian section and retaining only elliptic data. This establishes a broad correspondence between non-asymptotic spacetime regions and quantum states.

Assuming that $\mathbf{H}_\sigma$ factorizes over connected components of $\sigma$, our framework admits operators and partial traces.
This allows us to explore the information-theoretic structure of the states we define.
As an example, we construct a family of states by deforming partial Cauchy slices $\Sigma$ that straddle a two-sided black hole; $\sigma$ consists of two spheres.
We construct the reduced state on one sphere and find that its R\'enyi entropies are positive, monotonic, and sensitive to all aspects of $\Sigma$ and its complex deformation.
The von Neumann entropy, however, is controlled only by the maximin surface in the causal domain of $\Sigma$, independently of other parameters, so long as the complex deformation does not vanish. Our proposal may thus explain the efficacy of tensor network toy models of holography while transcending their limitations.
}

\makeatletter
\gdef\@fpheader{\mbox{}}
\makeatother

\begin{document}
\maketitle

\section{Introduction}

It would be of great interest to establish a direct correspondence between quantum states of gravity and arbitrary finite Lorentzian spacetime regions.
This would allow us to discuss quantum gravity regardless of asymptotic structure, and it would obviate the need for  Euclidean state preparation, which is cumbersome, unphysical, and restrictive.
Instead, we would be able to work directly with states that describe regions of the universe we live in. 

\subsection{Proposal}
\label{sec:intro1}

Here we propose the following correspondence.

\begin{figure}[tbp]
    \centering
    \includegraphics[width=1\linewidth]{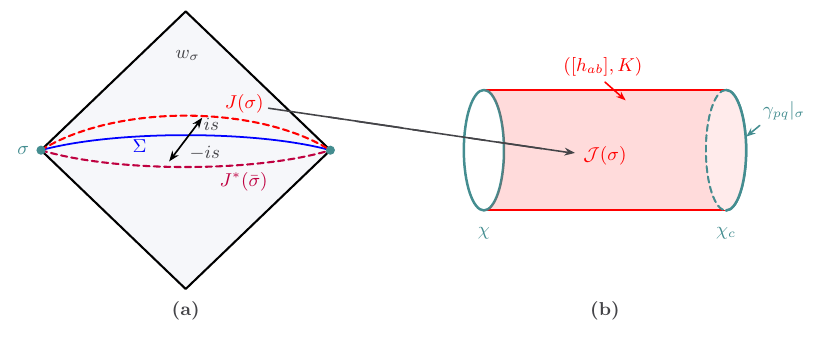}
    \begin{figurecaptionblock}
    \caption[A Lorentzian wedge and its elliptic data manifold]{Given a $(d+1)$-dimensional Lorentzian wedge $w_\sigma$, we construct a (non-unique) quantum state $\j(\sigma)$ that gives rise to $w_\sigma$ in the classical limit.
    (a) A $d$-dimensional Cauchy slice $\Sigma$ of $w_\sigma$ is deformed to a submanifold $J(\sigma)$ in the complexification of $w_\sigma$.
    In this example, the shared boundary $\sigma$ of $\Sigma$ and $J(\sigma)$ has two components, $\sigma=\chi\sqcup\chi_c$.
    (b) Of the full phase space data induced on $J(\sigma)$, we retain only the conformal data $([h_{ab}],K)$ in the interior of $J(\sigma)$, and the full $(d-1)$-metric $\gamma_{pq}|_\sigma$ on $\sigma$.
    These assignments turn $J(\sigma)$ into an elliptic data manifold $\mathcal J(\sigma)$. --- Conversely, we regard \emph{any} elliptic data manifold $\j(\sigma)$, with arbitrary complex $([h_{ab}],K)$ and real $\gamma_{pq}|_\sigma$, as a quantum state.
    Gluing $\j(\sigma)$ to its involution $\j^*(\bar\sigma)$ yields boundary conditions for the GPI; the output is the norm of $\j(\sigma)$.
    If the norm-GPI is dominated by a single saddle geometry, its CRT fixed point set is a Lorentzian wedge.
    It represents the classical limit of $\j(\sigma)$.}
    \label{fig:introwedge}
    \end{figurecaptionblock}
\end{figure}

\paragraph{Complex Cauchy-slice deformation} Let $w_\sigma$ be a $(d+1)$-dimensional Lorentzian spacetime wedge with finite edge $\sigma$; see Fig.~\ref{fig:introwedge}.
That is, $w_\sigma$ is a causal domain of dependence whose $d$-dimensional Cauchy slices all have the $(d-1)$-dimensional spatial surface $\sigma$ as their boundary.
Suppose that the classical metric and fields are real analytic functions of $d+1$ coordinates $(t,r,\ldots)$ on $w_\sigma$, so that we may view $w_\sigma$ as a complex manifold in a neighborhood of the real Lorentzian section.
Let $\Sigma$ be a Cauchy slice of $w_\sigma$, and suppose for simplicity that $\Sigma$ corresponds to $t=0$ and $\sigma$ to $r=1$.
Let $J(\sigma)$ be the embedded $d$-dimensional submanifold
\begin{equation}
    t = i s(r,\ldots),
\end{equation}
where $s$ is a real analytic function on $\Sigma$ that vanishes only at $r=1$.
Thus $\sigma$ is the boundary not only of $\Sigma$ but also of $J(\sigma)$.

\paragraph{Induced elliptic data}
Let $\{ h_{ab}, K_{ab} \}$ denote the induced metric and extrinsic curvature on $J(\sigma)$.
Let $\{([h_{ab}],K)_{J(\sigma)\setminus\sigma},(\gamma_{pq})_\sigma\}$ be the data subset consisting only of the conformal class of $h_{ab}$ and the trace of $K_{ab}$, in the interior of $J(\sigma)$, as well as the full induced $(d-1)$ metric $(\gamma_{pq})_\sigma$ on $\sigma$.
We shall refer to this data subset as \emph{mixed-conformal data}. The manifold
$J(\sigma)$, when equipped with mixed-conformal data, will be called an \emph{elliptic data manifold} and denoted by $\j(\sigma)$.

\paragraph{Space of elliptic data manifolds} We consider the vector space $\mathbf{V}_\sigma$ spanned by \emph{all} manifolds that have boundary $(\sigma,\gamma_{pq}|_\sigma)$ and are equipped with mixed-conformal elliptic data. Importantly, we do not restrict to $\j(\sigma)$ obtained by complex-deforming a Cauchy slice of a Lorentzian wedge, Eq.~\eqref{eq:deform}. 
We presented that procedure only to gain some intuition for how a classical-quantum correspondence might emerge. But the quantum state $\j(\sigma)$, not $w_\sigma$, is the primary object in our construction, and its mixed-conformal data can be specified freely. Whether $\j(\sigma)$ admits a classical approximation or classical limit is a secondary question to which we will return shortly.

\paragraph{Dual vectors, inner product, and norm} A dual vector or ``bra'' $\j^*(\bar\sigma)$ is obtained by reversing the orientation of $\j(\sigma)$ and complex conjugating its elliptic data. [Thus if $\j(\sigma)$ was obtained by Eq.~\eqref{eq:deform}, then the opposite deformation $t = -i s(r,\ldots)$ yields $\j^*(\bar\sigma)$.] A pre-inner product $\braket{\k(\sigma),\mathcal{L}(\sigma)}$ is defined by gluing $\k^*(\bar\sigma)$ to $\mathcal{L}(\sigma)$ across $\sigma$ and evaluating the GPI (over $(d+1)$-manifolds with complex metric) with the resulting closed boundary condition. In particular, gluing $\j(\sigma)$ to $\j^*(\bar\sigma)$ and evaluating the GPI computes the norm of $\j(\sigma)$. The pre-inner product is assumed to be positive semidefinite; quotienting by null states yields a Hilbert space with positive-definite inner product.

\paragraph{Norm geometry and CRT fixed locus} If the norm-GPI is dominated by a \emph{single} classical saddle geometry $(M,g_{ij})$ in the $G_N\to\ 0$ limit, then we call $(M, g_{ij})$ the \emph{norm geometry} of $\j(\sigma)$.
The (complexified) norm geometry is invariant under a CRT transformation (i.e., an antiholomorphic involution) $\boldsymbol{\Theta}$ that swaps $\j(\sigma)$ and $\j^*(\bar\sigma)$, or else $\boldsymbol{\Theta}(M,g_{ij})$ would furnish a second saddle with the same real part of the action. The fixed point set $\text{Fix}(\boldsymbol{\Theta})$ of $M$ under $\boldsymbol{\Theta}$ is a real $(d+1)$-manifold with real induced metric.

\paragraph{Classical spacetime limit} If the signature of the metric on $\text{Fix}(\boldsymbol{\Theta})$ is Lorentzian, then $w[\j(\sigma)]\equiv \text{Fix}(\boldsymbol{\Theta})$ is a Lorentzian wedge with edge $\sigma$. If in addition $\j(\sigma)$ can be smoothly deformed to a Cauchy slice $\Sigma[\j(\sigma)]$ of $w[\j(\sigma)]$, while preserving the appropriate subset of $(M,g_{ij})$ as the unique dominant saddle, then we call $w[\j(\sigma)]$ the \emph{classical spacetime limit} of the quantum state $\j(\sigma)$.

Several remarks are in order. If no norm geometry exists, or if the metric on its CRT fixed point set is not Lorentzian, or if the smooth-deformation criterion fails, then $w[\j(\sigma)]$ is not defined. In this case, $\j(\sigma)$ has no classical spacetime limit. We are not currently aware of any example of a mixed-conformal elliptic data manifold with this property. 
The classical limit criterion could be broadened to assign superpositions of classical geometries to ``sufficiently simple'' superpositions of states with classical limits, but we will not attempt this here.

\paragraph{Classical-quantum correspondence} If $\j(\sigma)$ is constructed by a sufficiently small complex deformation of a Cauchy slice of a given $w_\sigma$, as in Eq.~\eqref{eq:deform}, then it is interesting to ask whether $\j(\sigma)$ ``returns'' $w_\sigma$ as its classical spacetime limit:
\begin{equation}\label{eq:bi}
    w[\j(\sigma)]=w_\sigma.
\end{equation}
This relation does appear to hold broadly; we currently know of no counterexamples with $\sigma\neq\varnothing$ (except when Dirichlet data instead of mixed-conformal data are retained~\cite{shortpaper}).

However, Eq.~\eqref{eq:bi} does not hold universally. With $\j(\varnothing)$ constructed as a complex deformation of a slice $\Sigma$ of Lorentzian de Sitter space $w^{\rm dS}_\varnothing$, we find in Sec.~\ref{new:sec:lorentz} that $w[\j(\varnothing)]=\varnothing\neq w^{\rm dS}_\varnothing$. This may have some bearing on the puzzling Hilbert space structure of closed universes~\cite{Abdalla:2026nbp}; see also~\cite{Usatyuk:2024mzs,Usatyuk:2024isz,Harlow:2025pvj,Abdalla:2025gzn,Harlow:2026hky,Zhao:2026mpl,Nomura:2026igt}. It also raises the question of whether de Sitter space exists as the classical limit of any quantum state.  (Note that the existence of de Sitter space is not required from an empirical standpoint~\cite{DysonKlebanSusskind}.)


Given a Lorentzian wedge $w_\sigma$, let $\mathbf{C}(w_\sigma) = \{\j(\sigma)\in \mathbf{H}_\sigma: w[\j(\sigma)]=w_\sigma \}$ be the set of quantum states whose classical spacetime limit is $w_\sigma$. This defines a classical-quantum correspondence $w_\sigma \leftrightarrow \mathbf{C}(w_\sigma)$. Unless $\mathbf{C}(w_\sigma)$ is empty, as it appears to be for $w^{\rm dS}_\varnothing$, $\mathbf{C}(w_\sigma)$ will contain infinitely many states. We interpret this as a close analog of the choice of polarization when a quantum wavepacket is associated to a point particle with phase space data $(x,p)$. Concretely, we expect that increasing $s$ in Eq.~\eqref{eq:deform} increases the uncertainty in $([h_{ab}],K)$ around the values on $\Sigma$, and decreases the uncertainty in their canonical conjugates.

\begin{wrapfigure}{r}{0.4\textwidth}
    \vspace{-0.6\baselineskip}
    \noindent
    \includegraphics[width=0.30\textwidth]{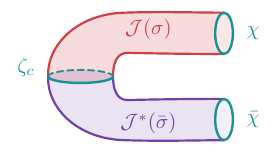}
    \begin{figurecaptionblock}
    \caption{A reduced state $\tilde \rho(\chi,\bar\chi)$ is constructed by gluing the state $\j(\sigma)$ to its involution $\j^*(\bar\sigma)$ across $\chi_c=\sigma\setminus\chi$, producing the joint $\zeta_c$.
    The surviving boundary is $\chi\sqcup\bar\chi$.}
    \label{fig:rho-chi-partial-gluing}
    \end{figurecaptionblock}
    \vspace{-0.5\baselineskip}
\end{wrapfigure}

\paragraph{Factorisation assumption} In order to test our proposal quantitatively, as we do in Sec.~\ref{sec:examples}, it is useful to introduce additional structure. We shall assume that $\mathbf{H}_\sigma$ factorises over the connected components of $\sigma$. For example, suppose that $\sigma = \chi \sqcup \chi_c$; then we assume that the Hilbert space is bipartite: $\mathbf H_\sigma=\mathbf H_\chi\otimes\mathbf H_{\chi_c}$.
With this assumption, partial traces can be defined and R\'enyi entropies computed. Given a state $\j(\sigma)$, a reduced state on $\chi$ is defined by the partial gluing of $\j(\sigma)$ to $\j^*(\bar\sigma)$ across $\chi_c$. This yields an elliptic data manifold whose boundary is $\chi\sqcup \bar\chi$. It represents the unnormalized density operator $\tilde \rho(\chi,\bar\chi)$; see Fig.~\ref{fig:rho-chi-partial-gluing}. The full gluing followed by GPI evaluation defines its trace, $\tr\tilde\rho(\chi,\bar\chi)$.
(The full trace agrees with the norm of $\j(\sigma)$ by construction, as it should.)

\begin{figure}[tbp]
\begin{minipage}{\textwidth}
    \centering
    \includegraphics[width=0.98\textwidth]{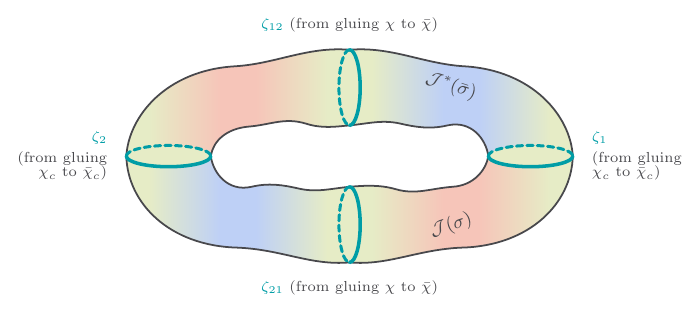}
\begin{figurecaptionblock}
   \caption{The closed elliptic data manifold $\tr_\chi[\tilde\rho(\chi,\bar\chi)^2]$ is obtained by cyclic gluing (here, across the vertical plane) of two copies of $\tilde \rho(\chi,\bar\chi)$ (see Fig.~\ref{fig:rho-chi-partial-gluing}).}
    \label{fig:rho2-cyclic-gluing}
\end{figurecaptionblock}
\end{minipage}
\end{figure}

Operator products are obtained by sequential gluing.
For example, $[\tilde\rho(\chi,\bar\chi)]^2$ is given by two copies of $\tilde \rho(\chi,\bar\chi)$, partially glued to one another.
Its trace is the two copies fully (cyclically) glued; and the trace is evaluated by the GPI; see Fig.~\ref{fig:rho2-cyclic-gluing}.
This allows us to compute the $n$-th R\'enyi entropy of the reduced state $\rho(\chi,\bar\chi)$ as 
\begin{equation}
    S_n[\rho(\chi,\bar\chi)]=\frac{1}{1-n} \log\frac{\tr[\tilde\rho(\chi,\bar\chi)^n] }{[\tr \tilde \rho(\chi,\bar\chi)]^n}.
\end{equation}
Analytic continuation to $n=1$ yields the von Neumann entropy $S_1$.

A simplified version of the above framework was proposed and examined in Ref.~\cite{shortpaper}. Some of the technical aspects of our proposal may evolve in response to further study of examples and applications.

\subsection{Relation to Previous Work}
\label{sec:previouswork}

\paragraph{Path integral approach to quantum gravity} Asymptotic applications of the GPI, including the seminal examples in Refs.~\cite{Gibbons:1976ue,Hawking:1982dh}, are special cases of our proposal.
Though often presented as computations of a partition function, they can equivalently be viewed as norms of states $\j(\sigma)$.
For example, a closed boundary of the form $S^1(\beta)\times S^{d-1}$ with real data can be cut in half to yield $\j(\sigma)$, $\j^*(\bar\sigma)$ of the form $[0,\beta/2]\times S^{d-1}$.

Our approach also encompasses closed universe amplitudes in quantum cosmology, as the special case $\sigma=\varnothing$.
For example, the unnormalized amplitude for the creation of a closed universe from nothing is an inner product $\braket{\k(\varnothing),\mathcal{L}(\varnothing)}$, where $\k$ is a closed $d$-manifold and $\mathcal{L}=\varnothing$ is the Hartle-Hawking state~\cite{Hartle:1983ai}.
(In a consistent application of the GPI, norms must also be computed from the GPI.
This implies that historical evaluations of the Hartle-Hawking proposal were inconsistent, as noted in Ref.~\cite{Abdalla:2026nbp}.)

\paragraph{AdS/CFT correspondence}

Our proposal is less than a nonperturbative quantum theory of gravity, but more than a formal Hilbert space construction. It is a tool for extracting quantum information theoretic properties of states of such a theory. We may hope that this tool will constrain, and help us construct, a theory that is directly applicable to the observed universe.

The assumption of Hilbert space factorisation is crucial for this aspect, as is our prescription for replicating and partially gluing elliptic data manifolds.
This set of ideas is rooted mainly in the role played by the GPI in the proof~\cite{Lewkowycz:2013nqa} of the Ryu-Takayanagi prescription~\cite{Ryu:2006bv} for computing the von Neumann entropy of boundary subregions in AdS/CFT.

It is important, however, to distinguish between state preparation in a known quantum theory such as the CFT, and our proposal that elliptic data manifolds represent states.
In AdS/CFT, the CFT state is the primary object in terms of which norms and R\'enyi entropies are defined, and a bulk-boundary dictionary~\cite{Gubser:1998bc,Witten:1998qj} must be invoked to determine the elliptic data for the GPI that computes the same quantities on the bulk side.
By contrast, we define states directly as elliptic data manifolds --- roughly, as half of a GPI boundary condition --- while remaining agnostic about the theory that governs those states.

For example, suppose that in some alternate history, the AdS/CFT correspondence had not been discovered by 2013.
Then the GPI ``proof''~\cite{Lewkowycz:2013nqa} of the Ryu-Takayanagi prescription~\cite{Ryu:2006bv} would instead have been interpreted as a GPI \emph{prediction} for the entanglement structure of the states of an \emph{unknown} quantum theory that lives, in some sense, on the conformal boundary of Anti-de Sitter space.
It would not have been difficult to recognize the entanglement entropies computed by the GPI as those of a CFT with a certain central charge.
This would not have told us precisely which CFT is the correct dual theory, but it would have provided a helpful insight nevertheless.
This alternate history is precisely the situation we find ourselves in when seeking a quantum gravity theory beyond AdS; and that is the setting in which our proposal may lead to some advances.

\paragraph{Tensor networks}

As mentioned above, state preparation in AdS/CFT is usually viewed as taking place in the CFT.
The corresponding boundaries for the gravitational path integral necessarily reside in an asymptotic region, usually of a Euclidean spacetime geometry.
An important change of perspective arrived with the introduction of MERA tensor networks~\cite{Vidal:2005zs,Vidal:2006qt} as toy models for the bulk-to-boundary map of AdS/CFT~\cite{Swingle:2009bg}.

Certain tensor networks~\cite{Pastawski:2015qua, Hayden:2016cfa} were able to model the RT prescription (with min-cuts playing the role of minimal area surfaces) and the quantum-error-correcting properties~\cite{Almheiri:2014lwa} of the bulk-to-boundary map.
In these contexts, tensor networks were viewed both as models for CFT state preparation, and as discretizations of Lorentzian slices of AdS.
Implicitly this supported a more direct notion of state preparation by Lorentzian slices.

\paragraph{Cauchy slice holography}

Building on a reinterpretation of tensor networks as Euclidean evolution~\cite{Caputa:2020fbc} in a $T\bar T$-deformed holographic CFT~\cite{McGough:2016lol, Hartman:2018tkw, Lewkowycz:2019xse, Gorbenko:2018oov, Belin:2020oib}, Ref.~\cite{Araujo-Regado:2022gvw} proposed that a $T\bar T$-deformed CFT lives on any Cauchy slice of an asymptotically locally AdS spacetime.
This ``Cauchy slice holography'' proposal is an important antecedent to our work, but our proposal differs significantly in several respects.

Our proposal is agnostic; its spirit is to deduce properties of quantum gravity from the properties of the states we define.
In particular, we do not invoke a specific quantum mechanical theory that lives on the elliptic data manifold $\j(\sigma)$.
As a result, we are able to operate in more general settings, both in that we do not require $\sigma$ to reside in an asymptotic region, and in that we need not base our construction on an asymptotically AdS spacetime in the first place.

Our $\j(\sigma)$ does not carry full initial value data but only elliptic data.
As such, it would not be meaningful to say a priori that an elliptic data manifold $\j(\sigma)$ is associated with a particular Cauchy slice of a particular Lorentzian geometry.
In our approach, however, it is critical that elliptic data are extracted from a given Lorentzian slice only after a small \emph{deformation} of its initial data \emph{away from the real Lorentzian sector}.
The complexified classical solution furnishes a well-defined saddle-point candidate in the GPI that computes the norm of $\j(\sigma)$.
The correspondence between $\j(\sigma)$ and its norm geometry defines the relation between quantum states and associated Lorentzian spacetime wedges.
To our knowledge, this aspect is entirely novel to our proposal.

Among further technical differences to Ref.~\cite{Araujo-Regado:2022gvw}, the most significant is our choice of mixed-conformal elliptic boundary conditions.
Dirichlet boundary conditions are not strongly elliptic~\cite{Anderson:2006bvp,Witten:2018lgb}, and the corresponding one-loop determinant need not therefore be well defined.
Moreover, Dirichlet boundary conditions have been shown to lead to negative norm states~\cite{Wall:2021bxi}.
As shown in Appendix~\ref{app:positivity_examples}, this problematic example does not survive with mixed-conformal boundary conditions.

\paragraph{Geometric constructions of states and operators}

Our geometric Hilbert space construction closely follows Marolf and Maxfield~\cite{Marolf:2020xie} (see also~\cite{Chen:2025fwp}); our construction of operators follows Ref.~\cite{MarolfColafranceschi}. These constructions have much older mathematical predecessors, especially the geometric assignment of state spaces and operators to boundaries and bordisms in the Atiyah--Segal formulation of topological and conformal field theory~\cite{Atiyah:1988,Segal:2004}.

In~\cite{Marolf:2020xie}, the closest recent work to ours, the focus is on the special cases where $\sigma=\varnothing$ (so that $\j(\sigma)$ is a closed universe or ``baby universe''~\cite{Marolf:2020xie}), or where $\j(\sigma)$ is asymptotically AdS in its norm geometry, so that $\j(\sigma)$ can be thought of as preparing a CFT state.
A common feature key to our work is the use of the GPI to evaluate inner products and full traces.
Here we consider more general elliptic data manifolds $\j(\sigma)$, whose boundaries need not be asymptotic; we establish a specific relation between \emph{deformed} Lorentzian data and quantum states; and we explore the substructure of the gravitational quantum states by computing the R\'enyi entropies of subsets of $\sigma$.

\paragraph{Beyond tensor networks}

In the family of examples studied in Sec.~\ref{sec:examples}, we will find that the von Neumann entropy of a boundary subregion $\chi\subset \sigma$ is given by the maximin surface~\cite{Wall:2012uf} homologous to $\chi$ in the norm geometry of $\j(\sigma)$, irrespective of all other parameters.
This relates to extant work in two ways.

The maximin result shows that our approach, despite its superficial similarity, transcends the tensor network framework.
The maximin surface does not lie on the Cauchy slice $\Sigma$ from whose deformation that elliptic data manifold $\j(\sigma)$ was obtained.
It is not the minimal surface on $\Sigma$.
A tensor network that discretizes $\Sigma$ cannot reproduce this result.
There is presently no satisfactory notion of a relativistic causal development of a tensor network --- i.e., there is no analogue of the Lorentzian spacetime wedge whose maximin surface defines the entropy.

The maximin result suggests, moreover, that the present approach may offer a path towards deriving a recent proposal for generalized entanglement wedges~\cite{Bousso:2022hlz, Bousso:2023sya}.
We will now turn to this and other interesting questions left for future work.

\subsection{Outlook}
\label{sec:outlook}

\paragraph{Generalized entanglement wedges}

As described in Sec.~\ref{sec:intro1}, let $w_\sigma$ be a Lorentzian spacetime wedge with boundary $\sigma$.
Let $\j(\sigma)$ be a quantum state whose complex norm geometry contains $w_\sigma$ as a fixed point set under CRT.
Let $\chi\subset\sigma$, and let $S_1(\chi)$ be the von Neumann entropy of the reduced state on $\chi$.

In the language of the present paper, the generalized entanglement wedge proposal~\cite{Bousso:2022hlz, Bousso:2023sya} becomes the prediction
\begin{equation}\label{eq:bpsa}
    S_1(\chi) = \frac{A_{\rm maximin}(\chi)}{4G_N}.
\end{equation}
Here $A_{\rm maximin}$ is the area of the maximin surface~\cite{Wall:2012uf} homologous to $\chi$ in $w_\sigma$.
(We have stated the classical limit for simplicity.) Sec.~\ref{sec:examples} confirms this prediction in a class of examples.

Conversely, our proposal furnishes a framework in which a proof of the proposal~\cite{Bousso:2022hlz, Bousso:2023sya} may be constructed. (Our $w_\sigma$ is the spacelike complement of the ``input region'' of~\cite{Bousso:2022hlz, Bousso:2023sya}; the latter will play no role in the proof. This distinguishes our approach from extant attempts~\cite{Balasubramanian:2023dpj, Kaya:2025vof} to prove the static limit~\cite{Bousso:2022hlz}.) Indeed, when the maximin surface lies in the interior of $w_\sigma$, the proof~\cite{Lewkowycz:2013nqa} of the maximin prescription in AdS/CFT~\cite{Ryu:2006bv,Hubeny:2007xt,Wall:2012uf} should generalize straightforwardly. In more general cases, the generalization would be nontrivial. We discuss this further at the beginning of Sec.~\ref{sec:examples}.

Qualitatively novel challenges arise because in our general setting, $w_\sigma$ is finite and $\sigma$ is not asymptotic.
Therefore, the maximin slice need not lie in the interior of $w_\sigma$.
A simple example of this arises when $w_\sigma$ is the causal domain of a shell in Minkowski or AdS.
We are currently analyzing examples of this type, along the lines of Sec.~\ref{sec:examples}.
This analysis may confirm Eq.~\eqref{eq:bpsa}, or it may necessitate a modification of Refs.~\cite{Bousso:2022hlz,Bousso:2023sya}.

Our discussion so far has somewhat oversimplified the proposal~\cite{Bousso:2022hlz, Bousso:2023sya}, which reduces to a maximin prescription only if max- and min-entanglement wedges of $\chi$, constructed from distinct prescriptions, coincide.
This can fail even in the classical limit, for example if $w_\sigma$ has a trapped edge component. The framework presented here could potentially explain this enigmatic feature, by demonstrating that replica-symmetry-breaking configurations $(M,g_{ij})$ contribute significantly~\cite{Akers:2019wxj} to the GPI that computes the entropies.

\paragraph{Positivity} A crucial assumption in our work is that the quantum states we define have positive definite norm. This is not manifest when the inner product is defined by a GPI. Counterexamples are known when Dirichlet boundary conditions are used~\cite{Wall:2021bxi}. The minimal assumption we need is that no such examples arise with \emph{some} suitable choice of elliptic data (perhaps, but not necessarily, the mixed-conformal boundary conditions used here) --- and in a class of gravity theories that should include, at the very least, General Relativity in 3+1 dimensions on the scales where it has been tested. As a first consistency check, we show in Appendix~\ref{app:positivity_examples} that with conformal or mixed-conformal boundary conditions, the negative norm saddle does not persist as a leading norm contribution. It will be of great interest to search for counterexamples to norm positivity with these and other boundary conditions. Extra conditions on $\j(\sigma)$ in a neighborhood of $\sigma$ may also be necessary in order to ensure positivity.

\paragraph{Nonperturbative completion}

The inner product, and hence the Hilbert spaces we construct, are defined only through the GPI. The GPI should be viewed as a perturbative asymptotic expansion about $G_N\to 0$, so Eq.~\eqref{eq-ZEinstein} need not converge. A complete construction of $\mathbf{H}_\sigma$ would require a nonperturbative definition of the inner product.
We must therefore regard the present construction as a GPI-approximation to an exact holographic Hilbert space in quantum gravity. This limitation is not unique to our work and is present in other GPI-based Hilbert-space definitions; see, for example, Ref.~\cite{Marolf:2020xie}. Alternatively, one may \emph{posit} a nonperturbative completion with suitable axiomatic properties, including the reality and positivity properties above. This is the perspective taken in Ref.~\cite{MarolfColafranceschi} in the context of asymptotically locally AdS boundary data.

\subsection{Outline}

In Sec.~\ref{sec:gpi}, we review the gravitational path integral and its saddle-point approximation, specializing to the case of mixed-conformal elliptic boundary data.
In Sec.~\ref{sec:formalism}, we construct a Hilbert space $\mathbf{H}_\sigma$ from $\mathbf{V}_{\!\sigma}$, the span of elliptic data manifolds $\j(\sigma)$.
A pre-inner product $\braket{\k(\sigma),\mathcal{L}(\sigma)}$ is defined on $\mathbf{V}_{\!\sigma}$ by gluing across $\sigma$ and evaluating the GPI with the resulting closed boundary condition.
Quotienting by null states and completion yields $\mathbf{H}_\sigma$.
In Sec.~\ref{new:sec:lorentz}, we discuss the correspondence between states $\j(\sigma)$ and Lorentzian spacetime wedges $w_\sigma$, already summarized above.

In Sec.~\ref{sec:sub}, we introduce additional structure by hypothesizing that $\mathbf{H}_\sigma$ is the tensor product of Hilbert spaces associated to individual connected components of $\sigma$.
With this assumption, elliptic data manifolds can be interpreted not only as bra and ket vectors, but also as operators and tensors.
Partial traces and operator products are defined by appropriate partial gluings; full traces are evaluated by the GPI.
In particular, we define the reduced density operator on a component $\chi$ of $\sigma$, and we explain how to compute its R\'enyi and von Neumann entropies.

In Sec.~\ref{sec:examples}, we illustrate all salient aspects of our proposal in an explicit example.
We consider a 3-parameter family of real analytic Lorentzian spacetime wedges $w_\sigma$.
For each member, we define a one-parameter family of associated quantum states $\j(\sigma)$ by complex deformation of a Cauchy slice $\Sigma$, via Eq.~\eqref{eq:deform}.
A reduced quantum state on a connected component $\chi\subset \sigma$ is constructed by partial gluing, as described above.
We find that its R\'enyi entropies are positive and monotonically decreasing, as required for consistency. Moreover, we find that the von Neumann entropy is universally set by the area of the maximin surface homologous to $\chi$ in $w_\sigma$.
This holds regardless of how far the maximin surface is from $\Sigma$, and regardless of other geometric features of $w_\sigma$ such as its size. Crucially, we find that the von Neumann entropy also does not depend on the strength of the complex deformation of $\Sigma$ to $\j(\sigma)$, \emph{so long as the deformation does not vanish}. When it does, we find that the von Neumann entropy jumps to a different value unrelated to the maximin surface of $w_\sigma$. This illustrates the critical importance of introducing a complex deformation of the real Cauchy slice in order to construct a quantum state that can have $w_\sigma$ as its unique classical spacetime limit.

Appendix~\ref{app:positivity_examples} discusses a violation of positivity exhibited by Wall~\cite{Wall:2021bxi}: certain real Dirichlet boundary conditions admit a pair of CRT-breaking saddle-point geometries; and no other saddle is known to dominate over this pair.
However, the conformal or mixed-conformal boundary data induced by these saddles are not CRT-symmetric and so do not correspond to a norm.
We resurrect the problem by constructing mixed-conformal boundary conditions that do represent a norm and admit a similar CRT-breaking pair of saddle points.
However, now an additional CRT-invariant saddle exists.
We find that it dominates and restores positivity.

\section{Gravitational Path Integral}
\label{sec:gpi}

Before defining a Hilbert space in Sec.~\ref{sec:formalism}, we first review the GPI and its saddle-point approximation.
In particular, we will introduce a \emph{mixed-conformal} boundary condition and its corresponding action.

\subsection{Definition}

The GPI is a map $G: \mathcal J(\varnothing) \to \mathbb{C}$, which can be formally written as
\begin{equation}\label{eq-GPI1}
G[\mathcal J(\varnothing)]
=\int_{(g_{ij},\varphi)\sim \mathcal J(\varnothing)} \frac{\mathcal{D} g_{ij}\,\mathcal{D}\varphi}{\mathrm{Diff}(g_{ij})}
\,e^{-I[g_{ij},\varphi]}.
\end{equation}
Here $\mathcal J(\varnothing)$ denotes a $d$-dimensional oriented closed smooth manifold with \emph{elliptic data}, i.e., data that can be imposed as a boundary condition on the boundary $\partial M$ of an oriented $(d+1)$-dimensional real manifold $M$ with complex metric $g_{ij}$ and fields $\varphi$.
The argument $\varnothing$ indicates that $\partial \mathcal J(\varnothing)$, the boundary of $\mathcal J(\varnothing)$, is empty.
$I[g_{ij}, \varphi]$ denotes the Euclidean action of the configuration (see subsection~\ref{subsec-Act}).
The path integral implicitly includes a sum over topologies of $M$.
We omit $\varphi$ in the remainder of this section for notational simplicity.

\subsection{Saddle-Point Approximation and Ellipticity}

The path integral~\eqref{eq-GPI1} can be evaluated exactly in certain toy models of gravity~\cite{Saad:2019lba}.
More generally, it can only be approximated by summing over saddle points in the limit $G_N \to 0$:
\begin{equation}\label{eq-ZEinstein}
G[\mathcal J(\varnothing)]\stackrel{G_N \to 0}{\approx} \sum_{g_{ij}}
\exp\!\left(-I[g_{ij}]\right)\,
G^{(1)}[g_{ij};\mathcal J(\varnothing)]\,
\Bigl(1 + O(G_N)\Bigr).
\end{equation}
Here the sum is over classical solutions $g_{ij}$ to the Einstein gravity \emph{boundary value problem} with boundary conditions $\mathcal J(\varnothing)$; and $G^{(1)}[g_{ij};\mathcal J(\varnothing)]$ denotes the one-loop determinant on the background $g_{ij}$.
The $O(G_N)$ term in parentheses denotes higher-loop corrections, and $\approx$ denotes an asymptotic expansion.

For Eq.~\eqref{eq-ZEinstein} to be well-defined, at least one solution must exist; we shall assume this.
Since Einstein gravity is non-renormalizable, computing $G[\mathcal J(\varnothing)]$ beyond one loop requires an appropriate UV completion whose existence we implicitly assume.
In practice, this will not matter for our leading-order analysis.
Moreover, 
the one-loop determinant $G^{(1)}[g_{ij};\mathcal J(\varnothing)]$ is well-defined only if the linearized gravity operator on the background $g_{ij}$ has finitely many zero modes.
This will be the case if the gauge-fixed boundary value problem is strongly elliptic.

For Einstein gravity, whether strong ellipticity holds depends on the type of boundary conditions that are specified.
In perturbative Euclidean gravity, conformal boundary conditions (see below) are elliptic in the sense relevant to one-loop quantization~\cite{Anderson:2006bvp,Witten:2018lgb}, whereas the more familiar Dirichlet boundary conditions (fixing the intrinsic metric of $\partial M$) are not elliptic.\footnote{Dirichlet boundary conditions are better behaved if $K_{ab}$ is positive or negative definite~\cite{Witten:2018lgb}, but $K_{ab}$ is not known \emph{a priori} at the nonperturbative level.}
We now turn to defining boundary conditions in detail.

\subsection{Mixed-Conformal Boundary Conditions}

Recall that $\mathcal J(\varnothing)$ denotes a $d$-dimensional closed oriented smooth manifold together with \emph{elliptic data} that furnish the boundary conditions on the configurations that the GPI sums over.

The elliptic data are ``half'' of the full phase space data induced on the boundary $\partial M$.
The full data consist of the intrinsic metric on $\partial M$, and the extrinsic curvature of $\partial M$ in $M$:
\begin{equation}\label{eq:bc_imposed1}
h_{ab} = e^i{}_a e^j{}_b\, g_{ij},
\qquad
K_{ab} = e^i{}_a e^j{}_b\,\nabla_{(i}n_{j)}.
\end{equation}
Here $e^i{}_a$ denote the tangential projectors from $M$ to $\partial M$~\cite{PoissonBook}, and $n_i$ is the unit conormal to $\partial M$: locally, if $\partial M$ is a level set of a function $\rho: M \to \mathbb{R}_{\geq 0}$ which vanishes on $\partial M$, then
\begin{equation}\label{eq-n1form}
    n_i=-\frac{\partial_i\rho}{\sqrt{g^{jk}\partial_j\rho\,\partial_k\rho}},
\end{equation}
where the branch of the square root is determined by the requirement that $n_i$ is outward pointing.

We now discuss which ``half'' of the full data should be specified.
A simple and intuitive choice would be Dirichlet boundary conditions, where one specifies all of $h_{ab}$ and no components of $K_{ab}$.
However, Dirichlet data do not furnish elliptic boundary conditions for Euclidean gravity~\cite{Anderson:2006bvp,Witten:2018lgb}, and a one-loop determinant need not be well-defined. Instead, we will propose a variant of conformal boundary conditions, so we begin by reviewing the latter~\cite{Coleman:2020jte, Banihashemi:2024flat, Banihashemi:2025thermal, Allameh:2025timelike}.

Conformal boundary conditions on $\partial M$ consist of the conformal class $[h_{ab}]$ of the intrinsic metric, along with the trace $K=K_{ab}h^{ab}$ of the extrinsic curvature:\footnote{The reader may wonder whether Hilbert-space constructions based on Dirichlet and (mixed-)conformal data yield the same Hilbert space, analogously to how definite-position and definite-momentum eigenstates in single-particle quantum mechanics are related by a change of basis. We are not aware of a mathematically precise argument for this in our case, and find it implausible in part because the failure of ellipticity suggests that a general Hilbert space construction based on Dirichlet data is impossible in the first place.}
\begin{equation}\label{eq:conformal_BC}
\mathcal J(\varnothing):([h_{ab}],\,K)_{J(\varnothing)},
\end{equation}
where $J(\varnothing)$ denotes the underlying manifold on which the elliptic data are placed.
The conformal class is the equivalence class of metrics under the identification $h_{ab} \sim e^{2\Omega} h_{ab}$, where $\Omega: J(\varnothing) \to \mathbb{C}$ is any smooth function.

\begin{figure}[tbp]
    \centering
    \includegraphics[width=0.86\textwidth]{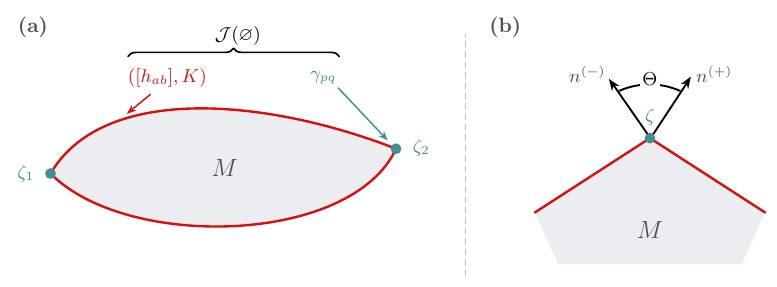}
    \begin{figurecaptionblock}
    \caption{GPI with mixed-conformal boundary data.
    (a) On a closed $d$-manifold $J(\varnothing)=\partial M$ with $(d-1)$-dimensional submanifold $\zeta$, we specify the full metric $\gamma_{pq}$ on $\zeta$ and $([h_{ab}],K)$ on $J(\varnothing)\setminus\zeta$.
    Together, $J(\varnothing)$ and these data comprise the closed elliptic data manifold $\mathcal J(\varnothing)$.
    In the example shown, $\zeta$ has two components, referred to as \emph{joints}.
    The GPI is over spacetimes $(M,g_{ij})$ that induce these data on their boundary.
(b) Local geometry at a joint $\zeta$.
The limiting outward normals $n^{(-)}$ and $n^{(+)}$ determine the exterior angle $\Theta$. The illustrated joint is convex and has $\Theta>0$.}
    \label{fig:mixed-conformal-data}
    \end{figurecaptionblock}
\end{figure}

Here we will use a small modification of conformal boundary conditions, which we call \emph{mixed-conformal}:
\begin{equation}\label{eq:mixed_conformal_BC}
\mathcal J(\varnothing):([h_{ab}],\,K)_{J(\varnothing)\setminus \zeta},\qquad \bigl(\gamma_{pq}\bigl)_{\zeta}.
\end{equation}
That is, we specify conformal data $([h_{ab}],K)$ on $J(\varnothing)$, \emph{except} on an embedded $(d-1)$-dimensional submanifold $\zeta$ of $J(\varnothing)$.
On $\zeta$, we shall specify Dirichlet data, i.e., the full $(d-1)$-dimensional metric $\gamma_{pq}$ (not just its conformal class); and correspondingly, we leave $K$ unspecified on $\zeta$.
See Fig.~\ref{fig:mixed-conformal-data}.
Note that $\zeta$ may be empty, and $\zeta$ may have multiple connected components,
\begin{equation}\label{eq-xis}
    \zeta = \zeta_1 \sqcup \ldots \sqcup \zeta_n,
\end{equation}
which we shall refer to as \emph{joints}.
We expect, though we do not prove, that generic mixed-conformal boundary conditions retain strong ellipticity, because they differ from the full conformal boundary conditions only on the joints, which are finite in number and measure-zero in volume.

Lastly, the orientation of $J(\varnothing)$ is a choice of equivalence class $[\omega_{J(\varnothing)}]$ of a nowhere-vanishing smooth top-form, where
\begin{equation}
\omega_{J(\varnothing)} \sim \omega_{J(\varnothing)}' \qquad \text{iff} \qquad \omega_{J(\varnothing)} = f\omega_{J(\varnothing)}'
\end{equation}
for some everywhere-positive smooth function $f$.
In the GPI, this equivalence class is imposed as a boundary condition by demanding that it agrees with the orientation induced on $\partial M$ by $M$, with the induced orientation defined by
\begin{equation}
    \omega_{M} = n \wedge \omega_{\partial M},
\end{equation}
where $n$ is the one-form defined in Eq.~\eqref{eq-n1form}.

In summary, an elliptic data manifold $\mathcal J(\varnothing)$ consists of a closed oriented manifold $J(\varnothing)$ (with orientation denoted by $[\omega_{J(\varnothing)}]$) together with the elliptic data in Eq.~\eqref{eq:mixed_conformal_BC}.

\subsection{Action}\label{subsec-Act}

We consider Einstein gravity with a cosmological constant of any sign. We do not expect the inclusion of matter fields or higher-curvature corrections to change our proposal substantially.
For the mixed-conformal boundary conditions described above, the action, including the appropriate boundary and joint terms, is \cite{Gibbons:1976ue,York:1972sj,Hayward:1993my}
\begin{align}
I[g_{ij}]
&=-\frac{1}{16\pi G_N}\int_M d^{d+1}x\,\sqrt{g}\,(R-2\Lambda)
-\frac{1}{8\pi G_N d}\int_{\partial M\setminus \zeta}d^d x\,\sqrt{h}\,K
\notag\\
&\hspace{2em}
-\frac{1}{8\pi G_N}\int_{\zeta}d^{d-1}x\,\sqrt{\gamma}\,\Theta.
\label{eq-IEinstein}
\end{align}
Here $R$ is the Ricci scalar and $\Lambda$ is the cosmological constant.
The second term is evaluated on the complement of the joints $\zeta$ in $\partial M$. 
The factor $1/d$ would be absent with Dirichlet boundary conditions but is appropriate since we fix conformal boundary conditions $([h_{ab}],K)$ on these portions of $\partial M$.
The final Hayward term arises from fixing the full intrinsic metric on $\zeta$.
The local geometry and angle convention are shown in Fig.~\ref{fig:mixed-conformal-data}(b).
At each point on $\zeta$, a signed exterior angle $\Theta$ is defined as follows: let $n^{(+)}_i$ and $n^{(-)}_j$ be the outward unit normals to the two smooth portions of $\partial M$ that meet at $\zeta$.
The exterior angle $\Theta$ is determined by
\begin{equation}
\cos\Theta=g^{ij}\,n^{(+)}_in^{(-)}_j,
\label{eq:dihedral_angle_def}
\end{equation}
with $\Theta$ chosen positive where the joint is convex from the point of view of $M$ and negative otherwise.

\section{Gravitational Hilbert Spaces}\label{sec:formalism}

Here, we define a holographic Hilbert space $\mathbf{H}_\sigma$ associated with any closed oriented $(d-1)$-dimensional manifold $\sigma$ that is equipped with a fixed intrinsic metric.
We will assume that $\mathbf{H}_\sigma$ factorises over the connected components of $\sigma$, and we will provide a geometric definition of a reduced state on a subset of connected components of $\sigma$.
Crucially, we will define traces and inner products using the GPI.

\subsection{Elliptic Data Manifolds With Boundary as Vectors}

\begin{figure}[tbp]
    \centering
    \includegraphics[width=0.75\textwidth]{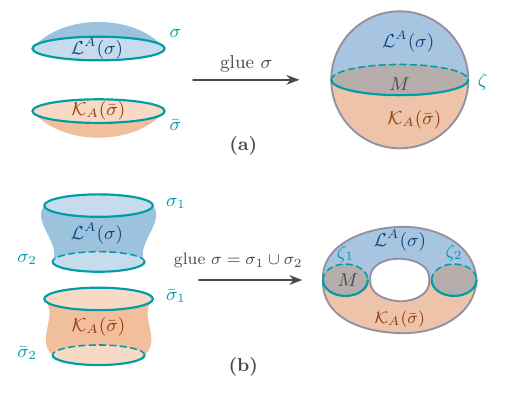}
    \begin{figurecaptionblock}
    \caption[Elliptic data manifolds with boundary and their gluing]{Elliptic data manifolds with boundary and their gluing.
    In panel (a), $\sigma$ is connected: the underlying manifolds $J(\sigma)$ and $K(\bar\sigma)$ are disks and glue across $\sigma$ to form a sphere.
    In panel (b), $\sigma=\sigma_1\sqcup\sigma_2$ has two connected components: the underlying manifolds are cylinders and glue along both circular boundaries to form a torus.
    Cyan marks the gluing loci, blue and orange mark the data-equipped elliptic data manifolds $\mathcal J^A(\sigma)$ and $\mathcal K_A(\bar\sigma)$, and the pale region shows one possible bulk filling $M$.
    The GPI sums over fillings.
    Setting $\mathcal K=\mathcal J^*$ gives $G[(\mathcal J^*)_A\mathcal J^A]=\langle\mathcal J,\mathcal J\rangle$.  }
    \label{fig:edm-gluing}
    \end{figurecaptionblock}
\end{figure}

We begin by extending the notion of an oriented \emph{closed} $d$-manifold [denoted $J(\varnothing)$ by itself, and $\j(\varnothing)$ when equipped with mixed-conformal data] to an oriented $d$-manifold \emph{with boundary}, denoted $J(\sigma)$ by itself and $\j(\sigma)$ when equipped with elliptic data.
See Fig.~\ref{fig:edm-gluing}.
The ``boundary'' $\sigma$ is technically an oriented $(d-1)$-dimensional Riemannian manifold that is mapped to the (actual topological) boundary $\partial J(\sigma)$ by a \emph{boundary parametrization}
\begin{equation}
    \phi_{J(\sigma)}: \sigma \to \partial J(\sigma),
\end{equation}
a diffeomorphism that preserves orientation; that is, the pullback $\phi^\star$ satisfies $[\omega_\sigma] = \phi^\star([\omega_{\partial J(\sigma)}])$.
Let $\zeta\supset \sigma$ be an oriented embedded submanifold of $J(\sigma)$.
Consistent with our convention for $J(\varnothing)$, we refer to the connected components of $\zeta\setminus\sigma$ as joints.
We define the elliptic data manifold $\j(\sigma)$ as $J(\sigma)$ together with mixed-conformal data:
\begin{equation}\label{eq:open_mixed_conformal_BC}
\j(\sigma):([h_{ab}],\,K)_{J(\sigma)\setminus \zeta},\qquad\bigl(\gamma_{pq}\bigl)_{\zeta}.
\end{equation}
Note that the $(d-1)$-dimensional metric $\gamma_{pq}$ on $\zeta$ must be specified on $\partial J(\sigma)$ in particular, and $K$ is not specified on $\partial J(\sigma)$.\footnote{Since a Hilbert space will be associated to the pair $(\sigma, \gamma_{pq}|_\sigma)$, it should strictly be denoted $\mathbf{H}_{\sigma,\gamma_{pq}|_\sigma}$.
We suppress the metric $\gamma_{pq}|_\sigma$ to lighten the notation.
The same remark applies to $\mathcal J(\sigma)$ itself, and to other objects indexed by $\sigma$ below, such as $\mathbf{S}_\sigma$, $\mathbf{V}_{\!
\sigma}$, $\mathbf{N}_\sigma$, etc.}

We now consider the set $\mathbf{S}_\sigma$ of all $\j(\sigma)$ with the same boundary $\sigma$, the same metric $\gamma_{pq}|_\sigma$ on $\sigma$, and the same orientation $[\omega_\sigma]$.
Let $\mathbf{V}_{\!
\sigma}$ be the free vector space generated by the set $\mathbf{S}_\sigma$, over the complex numbers:
\begin{equation}\label{eq-Vsigma}
\begin{aligned}
    \mathbf{V}_{\!\sigma}
    &= \textrm{span}\{\j^A(\sigma)\in\mathbf{S}_\sigma\} \\
    &\equiv \left\{ \sum_{k=1}^n c_k \j^A_k(\sigma):
    n<\infty,\ c_k\in\mathbb{C},\ \j^A_k(\sigma)\in\mathbf{S}_\sigma \right\}.
\end{aligned}
\end{equation}
We use abstract indices to keep track of the boundary ``slots'' of an elliptic data manifold when its tensorial role is relevant. In particular, we always display abstract indices in expressions that explicitly represent gluing or contraction. When referring to the same elliptic data manifold merely as a geometric object or quantum state, we suppress the abstract index.

For any $\sigma$, let $\bar{\sigma}$ denote the same manifold with opposite orientation $[\omega_{\bar{\sigma}}] = [-\omega_{\sigma}]$.
Let $\mathbf{S}_{\bar{\sigma}}$ be the set of all elliptic data manifolds $\k(\bar{\sigma})$, and let $\mathbf{V}_{\bar\sigma}$ be the free vector space generated by $\mathbf{S}_{\bar{\sigma}}$.
We interpret any element of $\mathbf{V}_{\bar\sigma}$ as a dual vector and denote it with a lower abstract index.
The action of a dual vector on a vector is the linear extension of the map 
\begin{equation}
    \k_A(\bar\sigma): \mathbf{V}_{\!
    \sigma} \to \mathbb{C},\qquad\j^A(\sigma) \mapsto G[\k_A(\bar\sigma)\j^A(\sigma)],
\end{equation}
where the repeated index $A$ denotes gluing.
Thus, the argument $\k_A(\bar\sigma)\j^A(\sigma)$ of the GPI is the closed elliptic data manifold obtained by gluing $\j(\sigma)$ to $\k(\bar\sigma)$ across $\sigma$; see Fig.~\ref{fig:edm-gluing}.\footnote{Explicitly, gluing is defined by identifying points on $\partial J(\sigma)$ with their image on $\partial K(\bar\sigma)$ under the diffeomorphism
$ \phi_{K(\bar\sigma)} \circ \iota \circ \phi_{J(\sigma)}^{-1}: \partial J(\sigma) \to \partial K(\bar\sigma)$.
Here $\iota: \sigma \to \bar\sigma$ is an orientation-reversing identity map on the underlying $\sigma$ manifold; its action ensures that the orientations of $\j(\sigma)$ and $\k(\bar\sigma)$ extend smoothly across the gluing locus, resulting in an oriented closed manifold.} 

\subsection{Hilbert Space Construction}

\paragraph{Involution and inner product}

The involution map
\begin{equation}\label{eq:inv}
    *:\mathbf{S}_\sigma \to \mathbf{S}_{\bar\sigma},\qquad\mathcal{J}(\sigma)\mapsto \mathcal{J}^*(\bar\sigma)
\end{equation}
acts on the elliptic data manifold $\mathcal{J}(\sigma)$ by complex conjugation of its data:
\begin{equation}
([h_{ab}],K)_{J(\sigma)\setminus\zeta} \mapsto (\overline{[h_{ab}]},\overline{K})_{J^*(\bar\sigma)\setminus\zeta},
\qquad
\gamma_{pq}|_\zeta \mapsto \gamma_{pq}|_\zeta,
\end{equation}
along with reversing its orientation, i.e. $[\omega_{J^*(\bar\sigma)}] = [-\omega_{J(\sigma)}]$.
Note that $\phi_{J^*} = \phi_J$ because the involution flips the orientation on both the manifold and its boundary, so the same boundary parametrization remains orientation-preserving.

We define an inner product on $\mathbf{V}_{\!
\sigma}$ by acting on a vector with a dual vector obtained by involution,
\begin{equation}
\begin{aligned}
    \braket{\cdot,\cdot}&:\mathbf{V}_{\!\sigma}\otimes\mathbf{V}_{\!\sigma}\to\mathbb{C},\\
    (\k(\sigma),\mathcal{L}(\sigma))
    &\mapsto \braket{\k(\sigma),\mathcal{L}(\sigma)}\\
    &\equiv G[\k^*_A(\bar\sigma)\mathcal{L}^A(\sigma)],
\end{aligned}
\end{equation}
and extending this map sesquilinearly.
(We remind the reader of the index conventions introduced earlier; in particular, the repeated index $A$ denotes gluing across the argument $\sigma$.)
By \eqref{eq-GPI1},
\begin{equation}
    G[\j^*(\varnothing)] = \overline{G[\j(\varnothing)]},
\end{equation}
which implies that the inner product has conjugate symmetry:
\begin{equation}
    \overline{\langle\mathcal{K}(\sigma), \mathcal{L}(\sigma)\rangle} = \langle\mathcal{L}(\sigma), \mathcal{K}(\sigma)\rangle.
\end{equation}

\paragraph{Null states, pre-Hilbert space, and completion} We \emph{assume} that the inner product is positive semidefinite: for any finite set of vectors $\{\mathcal{J}_m(\sigma)\}$ and complex coefficients $\{c_m\}$,
\begin{equation}\label{eq:RP_assumption}
\sum_{m,n}\overline{c_m}\,c_n\,
\langle\mathcal{J}_m(\sigma), \mathcal{J}_n(\sigma)\rangle \;\ge\; 0.
\end{equation}
We do not assume positive-definiteness, because related, less general GPI constructions are known to contain nonzero states with zero norm~\cite{Marolf:2020xie}.
Such states must be removed, by a standard process.
Define the null subspace
\begin{equation}
\mathbf{N}_\sigma
:=\left\{\,v\in\mathbf{V}_{\!
\sigma}\;:\;
\langle v, v\rangle=0\,\right\}.
\end{equation}
Under the positivity assumption \eqref{eq:RP_assumption}, $\mathbf{N}_\sigma$ is a subspace, and vectors in $\mathbf{N}_\sigma$ are orthogonal to all vectors.
We define a pre-Hilbert space by identifying vectors in $\mathbf{V}_{\!
\sigma}$ that differ by a null state:
\begin{equation}
\mathbf{H}^{(0)}_\sigma := \mathbf{V}_{\!
\sigma}\big/\mathbf{N}_\sigma.
\end{equation}
$\mathbf{H}^{(0)}_\sigma$ is equipped with the induced inner product, which will be positive definite by assumption \eqref{eq:RP_assumption}.
Finally, we define the holographic Hilbert space $\mathbf{H}_\sigma$ to be the completion of $\mathbf{H}^{(0)}_\sigma$ in the norm $\|v\| := \sqrt{\langle v, v\rangle}$.
The special case $\mathbf{H}_{\sigma = \varnothing}$ is the closed universe (or ``baby universe'') Hilbert space~\cite{Marolf:2020xie,Abdalla:2026nbp}.

In an abuse of notation, we will not distinguish between the vectors $\j^A(\sigma)\in \mathbf{V}_{\!
\sigma}$ and the elements of $\h_\sigma$, even though the latter are equivalence classes under the above quotient and should strictly be denoted $[\j^A(\sigma)]$.
The dual Hilbert space $\h^*_\sigma = \h_{\bar\sigma}$ is constructed analogously from $\mathbf{V}_{\bar\sigma}$.

\section{From Elliptic Data to Spacetime Regions and Back}
\label{new:sec:lorentz}

We have constructed a Hilbert space $\h_\sigma$, spanned by elliptic data manifolds $\j(\sigma)$ with boundary $\sigma$. In Sec.~\ref{sec:gtol}, we will construct a Lorentzian spacetime interpretation for some of these states. Conversely, in Sec.~\ref{sec:ltog}, we associate candidate quantum states to a given classical Lorentzian wedge.

\subsection{From a Gravitational State to a Lorentzian Wedge}
\label{sec:gtol}

Here we begin by asking whether a given state $\j(\sigma)$ has a classical spacetime limit $w[\j(\sigma)]$. We will propose necessary and sufficient conditions for its existence and an explicit construction of $w[\j(\sigma)]$ when they are met. We will then discuss some properties of the set of states with the same classical limit; and we will argue that the Cauchy slices of $w[\j(\sigma)]$ are inextendible at the semiclassical level, even when $\sigma\neq \varnothing$.

\paragraph{Norm geometry} Let $\j(\sigma)$ be a mixed-conformal elliptic data manifold. Recall that we use an upper abstract index to denote its role as a vector in Hilbert space, $\j^A(\sigma)$; that we reverse the orientation and complex conjugate the elliptic data of $\j^A(\sigma)$ to construct the corresponding bra-vector $\j^*_A(\bar\sigma)$; that the norm of $\j^A(\sigma)$ is represented by gluing the two manifolds across $\sigma$, yielding a closed elliptic data manifold  $\j^A(\sigma)\j^*_A(\bar\sigma)$; and that the norm is evaluated by computing the GPI with boundary condition $\j^A(\sigma)\j^*_A(\bar\sigma)$:
\begin{equation}\label{eq:norm}
    \braket{\j(\sigma), \j(\sigma)} = G[\j^*_A(\bar\sigma)\j^A(\sigma)].
\end{equation}
Suppose that the norm-GPI, Eq.~\eqref{eq:norm}, is dominated by exactly one leading saddle $(M, g_{ij})$. Then we call $(M, g_{ij})$ the \emph{norm geometry} of $\j(\sigma)$, and we interpret it as the classical geometry that most closely corresponds to $\j(\sigma)$. 

Note that $(M, g_{ij})$ is necessarily CRT invariant, i.e., its complexification $(M^{\mathbb{C}}, g_{ij})$ admits an antiholomorphic involution $\boldsymbol{\Theta}: M^{\mathbb{C}}\to M^{\mathbb{C}}$ that exchanges $\j^A(\sigma)$ with $\j^*_A(\bar\sigma)$ and thus preserves the elliptic data on $\j^A(\sigma)\j^*_A(\bar\sigma)$.
To see this, assume for contradiction that the complex manifold $\boldsymbol{\Theta}(M^{\mathbb{C}}, g_{ij})$ obtained by reversing the orientation of $M^{\mathbb{C}}$ and complex conjugating its metric and fields is distinct from $(M, g_{ij})$. However, $\boldsymbol{\Theta}(M^{\mathbb{C}}, g_{ij})$ would satisfy the same boundary condition $\j^A(\sigma)\j^*_A(\bar\sigma)$. It would thus contribute as a second saddle, with the same real part of the action. This contradicts the assumption that only one leading saddle exists.

\paragraph{Classical spacetime limit} The fixed point set (or ``fixed locus'') $\text{Fix}(\boldsymbol{\Theta})\subset M^{\mathbb{C}}$ defines a $(d+1)$-dimensional embedded real submanifold with real metric and field data. If the real metric of $\text{Fix}(\boldsymbol{\Theta})$ has Lorentzian signature,\footnote{Some of the conditions in this and the previous paragraph could be adjusted or refined in response to further exploration of this framework. For example, it is not clear that Lorentzian signature needs to be imposed as a condition; it may be sufficient to restrict allowed elliptic data by requiring, e.g., that the real part of $h_{ij}$ have Euclidean signature and the real part of $K$ be nonvanishing. Similarly, we expect dominance of a single saddle to be generic in a broad class of states. However, a possible alternative to the stated dominance condition would require only that $(M, g_{ij})$ dominate among \emph{saddles}. The smooth-approach condition should be either sharpened or omitted. We expect that explicit study of well-chosen examples will discriminate among these and other options.} then we define
\begin{equation}\label{eq:recovery}
    w[\j(\sigma)]\equiv \text{Fix}(\boldsymbol{\Theta}).
\end{equation}
Suppose, moreover, that $\j(\sigma)$ can be smoothly deformed towards (a Cauchy slice of) $w[\j(\sigma)]$ in $M^{\mathbb{C}}$, defining a foliation $\j_\eta(\sigma)$, $0<\eta\leq 1$; and suppose, finally, that the norm geometries of the states $\j_\eta(\sigma)$ are given by a nested family of subsets of $M^{\mathbb{C}}$. Then we regard the Lorentzian wedge $w[\j(\sigma)]$ as the \emph{classical spacetime limit} of $\j(\sigma)$.

\paragraph{Squeezed states} Given a Lorentzian wedge $w_\sigma$, let 
\begin{equation}
    \mathbf{C}(w_\sigma) = \{\j(\sigma)\in \mathbf{H}_\sigma: w[\j(\sigma)]=w_\sigma \}
\end{equation}
be the set of quantum states whose classical spacetime limit is $w_\sigma$. As noted in the introduction, this defines a classical-quantum correspondence $w_\sigma \leftrightarrow \mathbf{C}(w_\sigma)$. By the deformation condition articulated after Eq.~\eqref{eq:recovery}, $\mathbf{C}(w_\sigma)$ is either empty or has an infinite number of elements. We expect that $s$ in Eq.~\eqref{eq:deform} controls the tradeoff between the uncertainty in $([h_{ab}],K)$ and the uncertainty in their canonical conjugates around the classical initial data on $\Sigma$.\footnote{The elementary quantum-mechanical analogue is a squeezed Gaussian centered on a fixed classical phase-space point $(x_0,p_0)$,
$\psi_\Delta(x)\propto e^{-(x-x_0)^2/(4\Delta^2)+ip_0(x-x_0)/\hbar}
\propto e^{-(x-z_\Delta)^2/(4\Delta^2)}$, with
$z_\Delta=x_0+2i\Delta^2p_0/\hbar$.
Thus, at fixed $(x_0,p_0)$, changing the squeezing changes the displacement into complexified configuration space while leaving the classical state unchanged.} $\mathbf{C}(w_\sigma)$ can be made into a subspace of $\mathbf{H}_\sigma$ under a suitable definition of the classical limit of superpositions of elliptic data manifolds.

By definition, the elements of $\mathbf{C}(w_\sigma)$ all have the same classical limit. But they differ as quantum states. In particular, we expect their entanglement structure to differ when $\sigma$ admits a bipartition $\sigma=\chi\sqcup \chi_c$. Correspondingly, we expect that the von Neumann entropy of the reduced state on $\chi$ will be the same for all elements of $\mathbf{C}(w_\sigma)$; but the R\'enyi entropies may differ. Indeed, this is exactly the behavior we will find in the examples studied in Sec.~\ref{sec:examples}.

\paragraph{Classical limit as a firewall state} Classically, unless $\sigma=\varnothing$, $w[\j(\sigma)]$ can be extended to a spacetime wedge $w'[\j(\sigma)]\supsetneq w[\j(\sigma)]$ by adding extra initial data in a neighborhood of $\sigma$. But semiclassically, any smooth state in the larger spacetime wedge must have infinite entanglement across $\sigma$, and therefore the semiclassical state  obtained by reducing the state of $w'[\j(\sigma)]$ to $w[\j(\sigma)]$ will have divergent von Neumann entropy. It follows that $w[\j(\sigma)]$ cannot be so extended. In particular, modes of the quantum fields localized near $\sigma$ must be in an unentangled state similar to the Rindler vacuum \cite{Fulling:1973}. Such a state should have quantum singularities on the boundary of $w_\sigma$ \cite{Candelas:1977rindler}, for example in the expectation value of the stress-energy tensor, similar to a ``firewall'' state~\cite{Almheiri:2012rt}. 

\subsection{From a Lorentzian Wedge to a Gravitational State}
\label{sec:ltog}

In the previous subsection, we defined the set $\mathbf{C}(w_\sigma)$ of all states that have $w_\sigma$ as their classical spacetime limit. However, this definition was not constructive. Here we present a simple construction that associates \emph{candidate} quantum states to a given classical spacetime wedge $w_\sigma$. In large classes of examples, we expect that this actually yields elements  of $\mathbf{C}(w_\sigma)$. In de Sitter space, the construction fails; we give a preliminary argument (but do not prove) that $\mathbf{C}(w_\sigma)$ is empty in such cases.

\paragraph{Construction}
Let $(w_\sigma,g_{ij})$ be a real-analytic manifold with real-analytic Lorentzian metric $g_{ij}$ and compact Cauchy slice $\Sigma$; we will refer to $\sigma=\partial \Sigma$ as the \emph{edge} of $w_\sigma$. By allowing the coordinates to take complex values, $(w_\sigma,g_{ij})$ can be complexified. Real analyticity implies that the extension $w_\sigma\hookrightarrow w_\sigma^{\mathbb{C}}$ is holomorphic in an open neighborhood of $w_\sigma$.

For simplicity, assume that $\Sigma$ corresponds to $t=0$, $r<1$, so $\sigma$ is given by $t=0$, $r=1$. Let $s(r,\ldots)$ be a smooth function on $\Sigma$ that is strictly positive for $r<1$ and vanishes for $r=1$. If $s$ is chosen small enough to lie in the holomorphic extension of $w_\sigma$, then
\begin{equation}\label{eq:deform}
    t = i s(r,\ldots)
\end{equation}
defines a real $d$-dimensional embedded submanifold $J_s(\sigma)$ with boundary $\sigma$. Complex phase space data $h_{ab}, K_{ab}$ are induced on $J_s(\sigma)$ by $w_\sigma^{\mathbb{C}}$. We retain only the mixed-conformal data: $([h_{ab}], K)$ in the interior of $J_s(\sigma)$ and the induced metric $(\gamma_{pq})_\sigma$ on $\sigma$. This associates an elliptic data manifold, and thus a quantum state $\j_s(\sigma)\in\mathbf{H}_\sigma$, to $w_\sigma$.


\paragraph{Success criterion}
The above deformation protocol $w_\sigma\to \j_s(\sigma)$ will have succeeded if $\j_s(\sigma)\in \mathbf{C}(w_\sigma)$, i.e., if the resulting quantum state $\j_s(\sigma)$ has $w_\sigma$ as its classical spacetime limit: 
\begin{equation}\label{eq:twoway}
    w[\j_s(\sigma)]=w_\sigma,
\end{equation}
where $w[\j_s(\sigma)]$ is defined by Eq.~\eqref{eq:recovery}. In the family of examples studied in Sec.~\ref{sec:examples}, we find that the above success criterion is met.

But this is not guaranteed; in fact there are several failure modes. By construction, the original complex solution $(M,g_{ij})$ that contains $w_\sigma$ as a real Lorentzian section is a saddle-point candidate for the norm-GPI of $\j_s(\sigma)$. However, it is not obvious that the saddle $(M,g_{ij})$ lies on the correct integration contour. Even if it does, the saddle $(M,g_{ij})$ need not dominate; and even if it dominates, there may not exist a smooth deformation of $\j_s(\sigma)$ to $w[\j_s(\sigma)]$ as described after Eq.~\eqref{eq:recovery}. Let us exhibit an explicit example where Eq.~\eqref{eq:twoway} fails.

\paragraph{Failure example} Let $w^{\rm dS}_\varnothing$ be Lorentzian de Sitter space in 3+1 dimensions, with metric 
\begin{equation}
    ds^2=-dt^2 + \cosh^2 t d\Omega_3^2.
\end{equation}
Let $\Sigma = \{t=0\}$, so $\sigma=\varnothing$; and let $s\in (0,\pi/2)$. Then Eq.~\eqref{eq:deform} defines the state $\j_s(\varnothing)$ as a deformation of $\Sigma$. $\j_s(\varnothing)$ is a topological $S^3$ and carries the mixed-conformal data induced on $t=is$:
\begin{equation}
    [h_{ab}]=[d\Omega_3^2],\qquad K=-3 \tan s.
\end{equation}
The norm-GPI has boundary conditions consisting of the disjoint union $\j_s(\varnothing)$ and $\j_s^*(\varnothing)$. There are two $S^3$-symmetric saddles. The first saddle is the equatorial belt $-is \leq t \leq is$ of Euclidean de Sitter (the round $S^4$). This geometry contains $w^{\rm dS}_\varnothing$ as the Lorentzian CRT fixed locus $\operatorname{Im}(t)=0$. The second saddle consists of two disconnected geometries, each larger than half of Euclidean de Sitter: $-i\pi/2\leq t\leq is$ and $-is\leq t\leq i\pi/2$. CRT exchanges the two components so $\text{Fix}(\boldsymbol{\Theta})=\varnothing \neq w^{\rm dS}_\varnothing$. The second saddle has real Euclidean action $I_{\rm first}-\sqrt{3}\pi/G_N$, and thus it dominates over the first saddle.

\paragraph{Role of the smooth-deformation criterion} In the above example, the norm geometry of $\j_s(\sigma)$ fails to contain $w_\sigma$ regardless of the value of $s$. Suppose, however, that in some other example, we found that the norm geometry contains $w_\sigma$ but only for sufficiently large $s$. For these large $s$, it would be impossible to deform $\j_s(\sigma)$ to $w_\sigma$, or else we would have found small values of $s$ that succeed in preparing a quantum state with classical limit $w_\sigma$. Therefore large deformations, by themselves, cannot generate quantum states for $w_\sigma$ if small deformations cannot. One could consider eliminating the deformation criterion after Eq.~\eqref{eq:recovery}. This would be physically difficult to justify, at least if we interpret the $s\to 0$ limit as a squeezing limit as outlined above.  

\paragraph{Non-uniqueness}
If the assignment $w_\sigma\to\j_s(\sigma)$ succeeds, it will be far from unique. This immediately follows from the deformation criterion for the classical spacetime limit, and it is consistent with our discussion, in the previous subsection, of the cardinality of $\mathbf{C}(w_\sigma)$. Indeed, there is considerable freedom in the choice of a real function $s$.
Moreover,
we could choose any other Cauchy slice of $w_\sigma$ as our starting point.
All these different choices will in general define inequivalent quantum states on $\sigma$.

\paragraph{Non-degeneracy} Note that we required that the deformation $s$ be nonvanishing except on $\sigma$. This is important because otherwise the elliptic data extracted from $\j_s(\sigma)$ will be degenerate with the data on $\j_s^*(\bar\sigma)$. As a result, $\j_s(\sigma)$ will not ``remember'' $w_\sigma$.
As an extreme example, consider the case where $s$ vanishes everywhere.
Then the manifolds $\j_s(\sigma)$ and $\j_s^*(\bar\sigma)$ coincide with $\Sigma$ and have identical elliptic data, thus retaining only a subset of the initial data that would uniquely determine the wedge $w_\sigma$.
The same degenerate data are also consistent with different Lorentzian wedges.
This resembles the position eigenstate in quantum mechanics, for which the momentum (i.e., the other half of the elliptic data) is fully undetermined.

To explore this analogy further, consider a one-parameter family $\j_\eta(\sigma)$ of embedded submanifolds defined by setting
\begin{equation}\label{eq:deform-family}
    t = \pm i \eta s(r,\ldots),\qquad 0<\eta\leq 1,
\end{equation}
in the coordinate system described above.
As noted above, the original geometry $(M,g_{ij})$ is a candidate saddle in the GPI for the norm of $\j_\eta(\sigma)$, for all $\eta$.
But as $\eta\to 0$ for fixed $G_N$, the sliver of $M$ enclosed by the closed submanifold $(\j_\eta)^A(\sigma) (\j^*_\eta)_A(\bar\sigma)$ vanishes.
We expect quantum corrections to the saddle to diverge in this limit, as in the norm of a position eigenstate.

However, if $\eta$ is fixed as $G_N$ is taken to zero, then we may expect the GPI to be dominated by the classical saddle with lowest action.
This is the order of limits taken here; as a consequence, it suffices to require that $s$ vanish only on $\sigma$.

\section{Substructure of Gravitational States}
\label{sec:sub}

\subsection{Factorisation and Tensorial Structure}
\label{sec:factorisation_trace}

Recall that $\sigma$ may be composed of more than one connected component,
\begin{equation}\label{eq-factorised}
    \sigma = \sigma_1 \sqcup \ldots \sqcup \sigma_n.
\end{equation}
We \emph{posit} that the Hilbert space $\mathbf{H}_\sigma$ factorises over the components of $\sigma$:
\begin{equation}\label{eq:hfac}
   \mathbf{H}_\sigma = \mathbf{H}_{\sigma_1} \otimes \cdots \otimes \mathbf{H}_{\sigma_n}.
\end{equation}

By a trivial extension of the construction of vectors and covectors, denoted by $\j^A(\sigma)$ and $\j_A(\bar\sigma)$, respectively, we can construct tensors of different ranks.
For example, when $n=3$, we can define elliptic data manifolds that represent a $(3,0)$, $(2,1)$, $(1,2)$, or $(0,3)$ tensor.
We denote the specific role played by the data manifold by ``breaking up'' both the single index and the single argument attached to it, and using abstract tensor notation:
\begin{equation}
\begin{gathered}
    \mathcal{J}^{A_1 A_2 A_3}(\sigma_1, \sigma_2, \sigma_3),
    \qquad
    \mathcal{J}^{A_1 A_2}{}_{A_3}(\sigma_1, \sigma_2; \bar\sigma_3), \\
    \mathcal{J}^{A_1}{}_{A_2 A_3}(\sigma_1; \bar\sigma_2, \bar\sigma_3),
    \qquad
    \mathcal{J}_{A_1 A_2 A_3}(\bar\sigma_1, \bar\sigma_2, \bar\sigma_3).
\end{gathered}
\end{equation}
The semicolon separates the upper and lower boundary components in the argument.

The action of a tensor is defined, as before, by gluing and then evaluating the GPI with boundary conditions given by the resulting closed elliptic data manifold.
For example,
\begin{multline}
\mathcal{J}^{A_1 A_2}{}_{A_3}(\sigma_1, \sigma_2; \bar\sigma_3):
    \h^*_{\sigma_1}\otimes\h^*_{\sigma_2}\otimes\h_{\sigma_3} \to \mathbb{C}, \\
    (\j_{B_1}(\bar\sigma_1), \j_{B_2}(\bar\sigma_2), \j^{B_3}(\sigma_3)) \mapsto
    G[\j^{A_1 A_2}{}_{A_3}(\sigma_1, \sigma_2; \bar\sigma_3)\j_{A_1}(\bar\sigma_1) \j_{A_2}(\bar\sigma_2)\j^{A_3}(\sigma_3)].
\end{multline}
As before, repeated indices indicate gluing, i.e., the identification of the corresponding pair of boundary components $\sigma_i$.

A single abstract index may still label the entire boundary $\sigma$, even if it has multiple connected components. We split this index into separate indices for connected components only when those components are manipulated separately: for example, when some components are glued or traced while others remain as boundaries. Conversely, components that are treated together may share a single index; for example, $\j^A{}_B(\sigma_2;\bar\sigma_1\sqcup\bar\sigma_3)$ denotes a rank $(1,1)$ tensor.

As is standard, a $(k,\ell)$ tensor can be viewed not only as a map from $k$ factors of $\h^*_\sigma$ and $\ell$ factors of $\h_\sigma$ to $\mathbb{C}$, but more generally as an operator from any subset of these factors to the dual of the complementary factors.
For example,
\begin{multline}
\mathcal{J}^{A_1 A_2}{}_{A_3}(\sigma_1, \sigma_2; \bar\sigma_3):
    \h^*_{\sigma_1}\otimes\h^*_{\sigma_2} \to \h^*_{\sigma_3},  \\
    (\j_{B_1}(\bar\sigma_1), \j_{B_2}(\bar\sigma_2)) \mapsto
    \j^{A_1 A_2}{}_{A_3}(\sigma_1, \sigma_2; \bar\sigma_3)\j_{A_1}(\bar\sigma_1) \j_{A_2}(\bar\sigma_2),
\end{multline}
where the final expression denotes a manifold with boundary $\bar\sigma_3$, obtained by gluing across $\sigma_1$ and $\sigma_2$, viewed as an element of $\h^*_{\sigma_3}$.
Note the consistency of our index and repeated-index notation: only unrepeated indices survive as boundaries.
Note also that the former boundaries $\sigma_1$ and $\sigma_2$ become embedded submanifolds in the interior under gluing, so the final result is again a mixed-conformal elliptic data manifold.

The adjoint of an operator (defined as usual by $\braket{v,Ow} = \braket{O^\dagger v, w}$ for all $v,w$) is obtained straightforwardly by involution, Eq.~\eqref{eq:inv}, and exchange of upper and lower indices.
For example,
\begin{equation}\label{eq:adjoint}
    [\mathcal{J}^{A_1 A_2}{}_{A_3}(\sigma_1, \sigma_2; \bar\sigma_3)]^\dagger = (\mathcal{J}^*)^{A_3}{}_{A_1A_2}(\sigma_3; \bar\sigma_1, \bar\sigma_2).
\end{equation}
As usual, this construction extends linearly.

\subsection{Reduced Gravitational States and Entropy}
\label{sec:reduce}
\label{sec:renyi}

Consider an endomorphic operator on $\h_\sigma$:
\begin{equation}
    \mathcal{X}^A{}_B(\sigma;\bar\sigma):\mathbf{H}_\sigma\to\mathbf{H}_\sigma,\qquad
    \j^A(\sigma) \mapsto \mathcal{X}^A{}_B(\sigma;\bar\sigma) \j^B(\sigma),
\end{equation}
where as usual $\sigma$ may have more than one connected component.
We define its (full) trace by gluing the two copies of $\sigma$ and evaluating the GPI:
\begin{equation}\label{new:eq:geometric_full_trace}
    \tr_\sigma [\mathcal{X}^A{}_B(\sigma;\bar\sigma)]\equiv G[\mathcal{X}^A{}_A(\sigma;\bar\sigma)].
\end{equation}
Now consider any disconnected bipartition of $\sigma$,
\begin{equation}
    \sigma = \chi\sqcup\chi_c.
\end{equation}
(This loses no generality since $\chi$ and $\chi_c$ may each be empty or may contain more than one connected component.) By the above conventions, we can make this partition notationally explicit:
\begin{equation}
    \mathcal{X}^A{}_B(\sigma;\bar\sigma) = \mathcal{X}^{A_1A_2}{}_{B_1B_2}(\chi,\chi_c;\bar\chi,\bar\chi_c).
\end{equation}
We define the partial trace\footnote{A more general construction is needed to define a partial trace over subsets $\chi\subset\sigma$ with nonempty $(d-2)$-dimensional boundary $\partial\chi$: a $\delta$-neighborhood of $\partial\chi$ is excised from $J(\sigma)$.
In inner products and full traces, the resulting $(d-1)$-dimensional boundary is treated as the boundary of a tensionless end-of-the-world brane in $M$. Then one takes $\delta\to 0$. Details and examples will be presented elsewhere.} over $\chi_c$ (strictly, over $\h_{\chi_c}$) by
\begin{equation}
    \tr_{\chi_c}[\mathcal{X}^{A_1A_2}{}_{B_1B_2}(\chi,\chi_c;\bar\chi,\bar\chi_c)] =
    \mathcal{X}^{A_1A_2}{}_{B_1A_2}(\chi;\bar\chi);
\end{equation}
the result is an endomorphic operator on $\h_\chi$.


An operator $O$ on $\h_\sigma$ is called \emph{self-adjoint} if it satisfies $O=O^\dagger$; see Eq.~\eqref{eq:adjoint}.
A self-adjoint operator $\tilde \rho$ on $\h_\sigma$ is \emph{positive} if for all vectors $v\in \h_\sigma$, $\braket{v,\tilde \rho v}\geq 0$.
If in addition $\tr \tilde \rho\neq 0$, then we refer to $\tilde \rho$ as an \emph{unnormalized density operator}.
The corresponding normalized density operator is
\begin{equation}
    \rho =\frac{\tilde\rho}{\tr \tilde \rho};
\end{equation}
it satisfies $\tr \rho=1$.

A \emph{pure} density operator has a single nonzero eigenvalue.
Such an operator can be constructed from any elliptic data manifold $\j(\sigma)$ by duplication, involution of the duplicate, and assignment of one upper and one lower index:
\begin{equation}
    \tilde \rho^A{}_B(\sigma,\bar\sigma) = (\j^*)_B(\bar\sigma) \sqcup \j^A(\sigma).
\end{equation}
Note that the self-adjoint property is manifest.
Positivity follows from Eq.~\eqref{eq:RP_assumption}, also for linear extensions of this construction.

Upon bipartition $\sigma=\chi\sqcup\chi_c$, the partial trace of an unnormalized density operator $\tilde\rho(\sigma,\bar\sigma)$ defines the \emph{reduced} unnormalized density operator $\tilde\rho(\chi,\bar\chi)\equiv\tr_{\chi_c}\tilde\rho(\sigma,\bar\sigma)$ on $\h_\chi$, while preserving the norm.


The $n$-th R\'enyi entropy of an unnormalized density operator on $\h_\chi$ is given by
\begin{equation}
   S_n\bigl(\tilde \rho(\chi,\bar\chi)\bigr) \equiv \frac{1}{1-n} \log \frac{\tr_\chi[\tilde \rho(\chi,\bar\chi)^n]}{[\tr_\chi \tilde \rho(\chi,\bar\chi)]^n}.
\end{equation}
When $\tilde \rho(\chi,\bar\chi)$ has a geometric definition, this can be written as
\begin{equation}
\begin{aligned}
   S_n\bigl(\tilde\rho(\chi,\bar\chi)\bigr)
   &= \frac{1}{1-n}\Bigl[\log G\Bigl(
   \\[-2pt]
   &\qquad
   \tilde\rho^{A_1}{}_{A_2}(\chi,\bar\chi)\,
   \tilde\rho^{A_2}{}_{A_3}(\chi,\bar\chi)\cdots
   \tilde\rho^{A_n}{}_{A_1}(\chi,\bar\chi)\Bigr)
   \\[-2pt]
   &\qquad{}-n\log G\Bigl(\tilde\rho^{A}{}_{A}(\chi,\bar\chi)\Bigr)\Bigr].
\end{aligned}
\end{equation}
Note that the first GPI's boundary condition is obtained by cyclic gluing of $n$ copies of the elliptic data manifold that prepares $\tilde \rho(\chi,\bar\chi)$.
For $n=2$, this gluing is illustrated in Fig.~\ref{fig:rho2-cyclic-gluing}.

The von Neumann entropy $S_1$ of $\tilde\rho(\chi,\bar\chi)$ is defined by analytic continuation:
\begin{equation}
    S_1 = \lim_{n\to 1} S_n\bigl(\tilde \rho(\chi,\bar\chi)\bigr).
\end{equation}


\section{Example: Entropy of a Reduced Gravitational State}
\label{sec:examples}

In this section, we study the substructure of a family $(r_1,r_a,R;\epsilon)$ of gravitational states in 2+1-dimensional gravity with negative cosmological constant. (We choose $\Lambda<0$ in order to be able to work with 2+1-dimensional black holes; our analysis makes no use of the asymptotically AdS structure.)
Our starting point is a maximal-volume partial Cauchy slice $\Sigma$ that passes through the interior of a nonrotating BTZ black hole \cite{Banados:1992wn}  with radius $r_1$; see Fig.~\ref{fig:deflong-btz-penrose}. Such slices are parametrized by the minimum radius $r_a<r_1$ they reach inside the black hole. The third and final parameter that fixes $\Sigma$ is the radius $R$ of its finite boundaries $\sigma=\chi\sqcup\chi_c$ in the left and right exteriors, $R>r_1$.

The fourth parameter, $\epsilon>0$, is the overall strength of a complex deformation of $\Sigma$ that defines a mixed-conformal elliptic data manifold (EDM) $\mathcal J(\sigma)$ via Eq.~\eqref{eq:deform} (see Sec.~\ref{sec:ltog} for details).
$\mathcal J(\sigma)$ represents a pure state in $\mathbf{H}_\sigma=\mathbf{H}_\chi\otimes\mathbf{H}_{\chi_c}$.

We will compute the R\'enyi and von Neumann entropies of the reduced state on $\mathbf{H}_\chi$ at leading order in $G_N$, using the prescription of Sec.~\ref{sec:reduce}. The R\'enyi entropies will be sensitive to all four parameters; however, we will find that the von Neumann entropy depends only on $r_1$:
\begin{equation}\label{eq:s1}
    S_1[\rho(\chi,\bar\chi)]=\frac{2\pi r_1}{4G_N}.
\end{equation}


This result is noteworthy for two reasons: its value, and its universality. The value of $S_1$ agrees with the conjecture that $S_1$ should be the Bekenstein-Hawking entropy of the maximin surface homologous to $\chi$~\cite{Bousso:2023sya}. (The significance and implications of this agreement were discussed in Sections~\ref{sec:previouswork} and \ref{sec:outlook}.) This agreement requires, of course, that $S_1$ be universal, i.e., that it must not depend on the solution parameters $(R, r_a)$, nor on the deformation strength $\epsilon$.

It is worth examining the universality of $S_1$ more carefully. The independence of $S_1$ from $(R, r_a)$ can be understood in terms of the $n\to 1$ limit of $\mathbb{Z}_n$-quotiented saddles, which makes manifest that the von Neumann entropy depends only on the fixed-point set of the quotient~\cite{Lewkowycz:2013nqa}. Indeed, subject to the usual assumption of replica symmetry of the dominant saddle, we expect that a straightforward generalization of Ref.~\cite{Lewkowycz:2013nqa} will furnish a proof of the generalized entanglement wedge proposal~\cite{Bousso:2022hlz,Bousso:2023sya}, in those settings where the maximin surface homologous to a connected component $\chi\subset \sigma$ resides in the interior of $w[\j(\sigma)]$. (This will be the case whenever $\sigma$ has positive outward null expansions in $w[\j(\sigma)]$.)

The independence of $S_1$ from $\epsilon$ is quite subtle: it holds only for $\epsilon>0$. There is a phase transition at $\epsilon=0$; see Fig.~\ref{fig:deflong-renyi-vs-n}. This is a consequence of the fact that $\j(\sigma)$ has the same classical spacetime limit\footnote{This does not mean that the von Neumann entropy for $\epsilon=0$ has no geometric interpretation at all: one finds that it equals the entropy of a BTZ black hole into which the elliptic data embed as time-reflection symmetric data. We have not explored the physical significance of this observation.} (in the sense defined in Sec.~\ref{sec:gtol}), for all $\epsilon>0$; but it does not have such a limit for $\epsilon=0$. At $\epsilon=0$, the complexified Lorentzian solution $(r_1,r_a,R)$ degenerates as a filling for the GPI, because $\j(\sigma)$ coincides with  $\j^*(\bar\sigma)$ in it. Conversely, the degenerate filling cannot be complexified to a solution $(M^{\mathbb{C}},g_{ij})$ whose CRT fixed point set would define a unique classical spacetime limit $w[\j(\sigma)]$.

\subsection{Extracting Elliptic Data From a Deformed Finite Cauchy Slice}
\label{subsec:deformed-long-slice}

{ We first specify the real Lorentzian slice whose complex deformation will define the EDM preparing the state that we will be interested in.}
We work in the nonrotating BTZ geometry
\begin{equation}
 ds^2=-(r^2-r_1^2)dt^2+\frac{dr^2}{r^2-r_1^2}+r^2d\phi^2,
 \qquad \phi\sim\phi+2\pi,
 \label{eq:deflong-btz-metric}
\end{equation}
where $r_1$ is the horizon radius.

\begin{figure}[tbp]
    \begin{minipage}{\linewidth}
      \centering
      \includegraphics[width=0.5\linewidth]{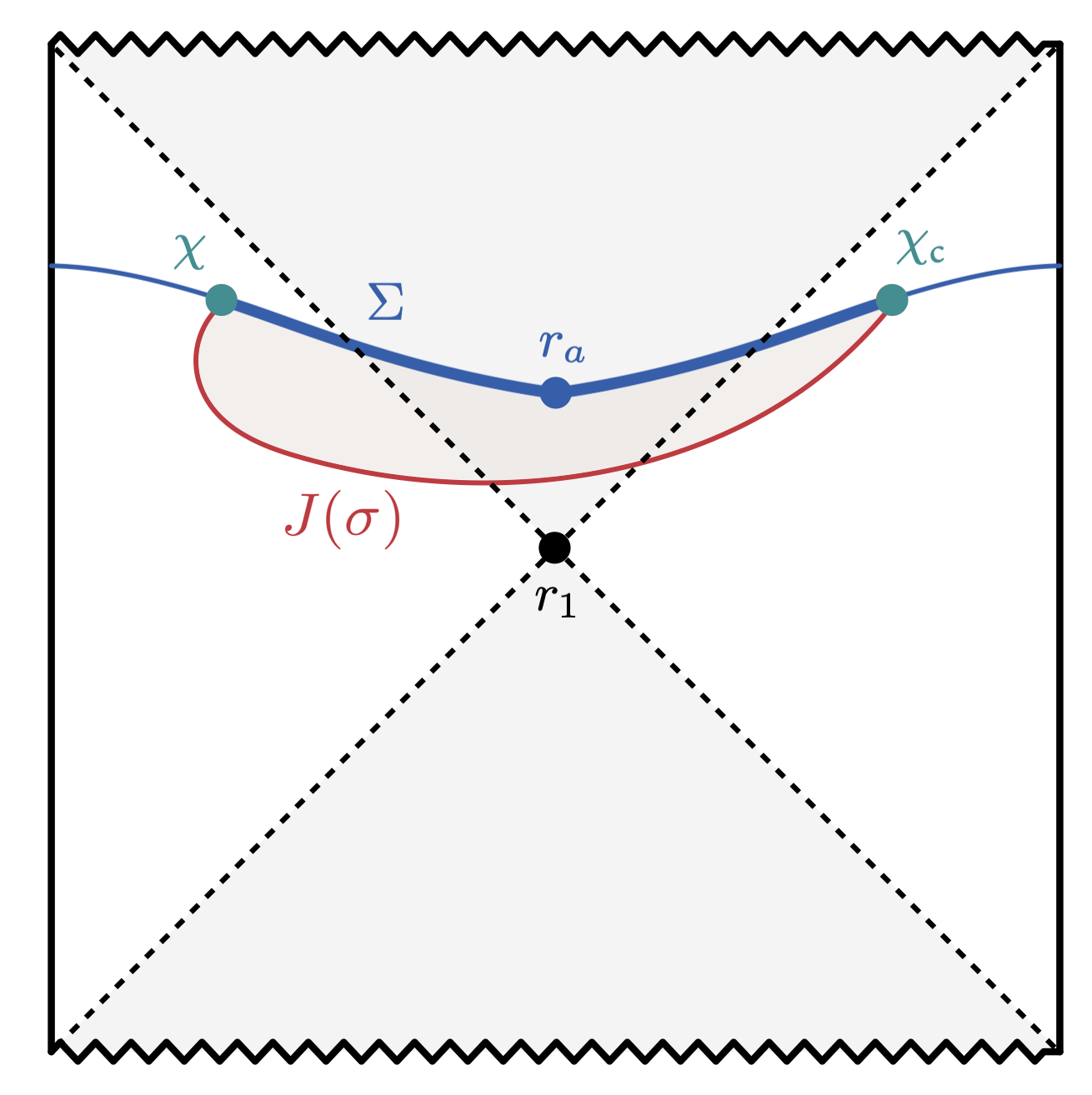}
      \begin{figurecaptionblock}
        \caption[A maximal slice and its finite-cutoff deformation]{
        BTZ black hole of radius $r_1$.
        A global maximal-volume slice (thin blue) with minimum radius $r_a<r_1$ contains the partial Cauchy slice $\Sigma$ (thick blue), which is bounded by circles $\chi$ and $\chi_c$ at $r=R$ in the left and right exteriors.
        A deformation of $\Sigma$ off the real Lorentzian sector yields the underlying preparation manifold $J(\sigma)$, with boundary $\sigma = \chi\sqcup \chi_c$.
        Together with the induced mixed-conformal data, it defines the EDM $\mathcal J(\sigma)$.}
        \label{fig:deflong-btz-penrose}
        \label{fig:deflong-complex-deformation}
      \end{figurecaptionblock}
    \end{minipage}
\end{figure}

For simplicity and definiteness, we take $\Sigma$ to be part of a left--right-symmetric maximal volume slice embedded in a BTZ black hole.
Specifying such a partial Cauchy slice requires the corresponding BH horizon radius $r_1$, the partial Cauchy slice cutoff radius $R>r_1$ and its minimum radius $r_a<r_1$.
The intrinsic metric and extrinsic curvature are\footnote{We use the convention of Eqs.~\eqref{eq:bc_imposed1} and \eqref{eq-n1form}, in which the normal is algebraically normalized by $n^\mu n_\mu=1$.
When this definition is continued to the real Lorentzian slice, the normal, and hence the extrinsic curvature, is purely imaginary.}
\begin{equation}
 ds_{\Sigma,0}^2
 =\frac{r^2dr^2}{r^4-r_1^2r^2+\mathcal E_0^2}+r^2d\phi^2,
 \qquad
 K^r{}_r=-K^\phi{}_{\phi}=-\frac{i\mathcal E_0}{r^2},
 \label{eq:deflong-undeformed-data}
\end{equation}
where we define $\mathcal E_0=r_a \sqrt{r_1^2-r_a^2}$, and the coordinate ranges are $\phi\in[0,2\pi)$ and $r\in [r_a,R]$; see Fig.~\ref{fig:deflong-btz-penrose}.

To describe the deformation, we introduce a coordinate $u$ for later convenience,
\begin{equation}
 r(u)^2=r_a^2\cosh^2u-(r_1^2-r_a^2)\sinh^2u,
 \qquad
 ds_{\Sigma,0}^2=du^2+r(u)^2d\phi^2.
 \label{eq:deflong-u-r-main}
\end{equation}
Note that for our left--right symmetric slice, $u\in [-u_R,u_R]$ with $r(u_R)=R$.
We then introduce Gaussian-normal coordinates $(\delta,u,\phi)$, where $\delta=0$ on the slice and $\delta$ is the proper time along the timelike geodesics orthogonal to the slice, so that the spacetime metric is
\begin{align}
 ds^2={}&-d\delta^2
 +\left(\cos\delta-\sin\delta\,\frac{\mathcal E_0}{r(u)^2}\right)^2du^2
 \notag\\
 &+r(u)^2\left(\cos\delta
 +\sin\delta\,\frac{\mathcal E_0}{r(u)^2}\right)^2d\phi^2.
 \label{eq:deflong-gnc}
\end{align}
The radius function is then $r(\delta,u)^2=[r(u)^2 \cos\delta+\mathcal E_0\sin\delta]^2/r(u)^2$.
Choosing a $\delta=\delta(u,\phi)$ specifies a deformation.
A simple constant imaginary displacement $\delta=i\epsilon$, with real $\epsilon$, gives an exact deformation in the interior, but it does not end on the prescribed real cutoff circles.
We therefore choose a real profile $\epsilon_R(u)$ that vanishes at $u=\pm u_R$, so that the deformed slice returns to the undeformed real section at $r=R$:\footnote{For the numerical calculation, we use $\epsilon_R(u)=\epsilon[1-(u/u_R)^p]^q$ with $p=8$ and $q=3$.
This profile is nearly constant in the interior and obeys $\epsilon_R'(\pm u_R)=0$.
The vanishing endpoint derivative is a convenient profile choice only.}
\begin{equation}
 \delta=i\epsilon_R(u),
 \qquad
 \epsilon_R(\pm u_R)=0,
 \qquad
 \epsilon_R(u)\simeq\epsilon\quad (|u|\ll u_R).
 \label{eq:deflong-profile-graph}
\end{equation}

For this deformed slice we read off the mixed-conformal data: we first compute the intrinsic geometry of the deformed surface.
Pulling back the Gaussian-normal metric to $\delta=i\epsilon_R(u)$ gives
\begin{align}
 h_{ab}dx^a dx^b={}&\left[(\epsilon_R')^2+
 \left(\cosh\epsilon_R-i\sinh\epsilon_R
 \frac{\mathcal E_0}{r(u)^2}\right)^2\right]du^2
 \notag\\
 &+r(u)^2\left(\cosh\epsilon_R+i\sinh\epsilon_R
 \frac{\mathcal E_0}{r(u)^2}\right)^2d\phi^2,
 \label{eq:deflong-profile-metric}
\end{align}
and we denote the circumference radius of the deformed surface by
\[
 r_0(u)=\sqrt{h_{\phi\phi}}
 =r(u)\left(\cosh\epsilon_R(u)+i\sinh\epsilon_R(u)\frac{\mathcal E_0}{r(u)^2}\right).
\]
Second, we determine the conformal class. 
Since our slice has the topology of a two-dimensional cylinder, the uniformization theorem implies that its induced metric is always conformally equivalent to the flat metric $dx^2 + d\phi^2$.
The only distinguishing feature is the normalized modulus. 
For our metric it is 
\begin{equation}
 m_R=\frac{1}{2\pi}\int_{-u_R}^{u_R}du\,
 \frac{\sqrt{(\epsilon_R')^2+
 \left(\cosh\epsilon_R-i\sinh\epsilon_R
 \mathcal E_0/r(u)^2\right)^2}}
 {r(u)\left(\cosh\epsilon_R+i\sinh\epsilon_R
 \mathcal E_0/r(u)^2\right)},
 \label{eq:deflong-modulus-target}
\end{equation}
where the square-root branch is continued from $\epsilon_R=0$ so that $\operatorname{Re}(m_R)>0$.
$u$ and the conformal coordinate $x$ are related by  $dx/du=\sqrt{h_{uu}(u)}/r_0(u)$ and we set $x(0)=0$, so that $x(\pm u_R)=\pm\pi m_R$.

The third piece of mixed-conformal data is the trace $K$ of the extrinsic curvature.
For the deformed slice, the unit normal vector has components
\begin{align*}
 n^\delta(\delta,u)&=
 i\frac{\cos\delta-\mathcal E_0\sin\delta/r(u)^2}
 {\sqrt{\left(\cos\delta-\mathcal E_0\sin\delta/r(u)^2\right)^2+\epsilon_R'(u)^2}},
 \\
 n^u(\delta,u)&=
 -\frac{\epsilon_R'(u)}
 {\left(\cos\delta-\mathcal E_0\sin\delta/r(u)^2\right)
 \sqrt{\left(\cos\delta-\mathcal E_0\sin\delta/r(u)^2\right)^2+\epsilon_R'(u)^2}},
 \\
 n^\phi&=0.
\end{align*}
The extrinsic curvature is
\begin{align}
 K(u)&=
 \left.\frac{1}{\sqrt{-g}}\left[
 \partial_\delta\left(\sqrt{-g}\,n^\delta\right)
 +\partial_u\left(\sqrt{-g}\,n^u\right)
 \right]\right|_{\delta=i\epsilon_R(u)},
 \notag\\
 \sqrt{-g}&=
 r(u)\left(\cos^2\delta-\frac{\mathcal E_0^2}{r(u)^4}\sin^2\delta\right).
 \label{eq:deflong-profile-trace}
\end{align}
The derivatives in \eqref{eq:deflong-profile-trace} are taken before setting $\delta=i\epsilon_R(u)$.
This determines $K(u)$ explicitly from the chosen profile.
As a function of the marked conformal coordinate, the prescribed trace is $K(x)=K(u(x))$.\footnote{
At fixed $R$, different deformation profiles can give different functions $K(u)$, and hence different functions $K(x)$ and different Hayward angles.
Our convenient choice $\epsilon_R'(\pm u_R)=0$ makes the Hayward angle real even though the metric can be complex.}

We can now extract the mixed-conformal EDM from the induced data of the deformed Cauchy slice. 
In the notation of Sec.~\ref{sec:factorisation_trace},
\begin{equation}
 \mathcal J^{AB}(\chi,\chi_c):\quad
 \left\{
 \begin{aligned}
 &\bigl([h_{ab}],K(x)\bigr)
 &&\text{on }\operatorname{int}J(\sigma),\\
 &\gamma_{pq}dx^p dx^q=R^2d\phi^2
 &&\text{on }\chi\sqcup\chi_c.
 \end{aligned}
 \right.
 \label{eq:deflong-conformal-data}
\end{equation}
Here the indices $A$ and $B$ label the unglued boundaries $\chi$ and $\chi_c$, respectively.
The first line fixes the conformal class and the trace in the interior of $J(\sigma)$, while the second fixes the full real metric on its two boundary circles.
The involuted EDM $(\mathcal J^*)_{AB}(\bar\chi,\bar\chi_c)$ carries the complex-conjugate data and the opposite orientation.

The union of $\mathcal J^{AB}(\chi,\chi_c)$ and $(\mathcal J^*)_{AB}(\bar\chi,\bar\chi_c)$ (without any gluings) represents an unnormalized pure state density operator on $\mathbf{H}_\chi\otimes\mathbf{H}_{\chi_c}$.
Its partial trace over $\mathbf{H}_{\chi_c}$ is the reduced density operator $\tilde\rho(\chi,\bar\chi)$, with components
\begin{equation}
 \tilde\rho^A{}_{A'}(\chi,\bar\chi)
 =(\mathcal J^*)_{A'B}(\bar\chi,\bar\chi_c)\,
 \mathcal J^{AB}(\chi,\chi_c),
 \label{eq:deflong-reduced-density-operator}
\end{equation}
where the repeated index $B$ denotes gluing across $\chi_c$.
The cyclic gluing that computes the $n$-fold trace is the closed EDM
\begin{equation}
 \mathcal J_n(\varnothing)
 =\prod_{i=1}^{n}\left[
 (\mathcal J^*)_{A_{i+1}B_i}(\bar\chi,\bar\chi_c)\,
 \mathcal J^{A_iB_i}(\chi,\chi_c)
 \right],
 \qquad A_{n+1}\equiv A_1.
 \label{eq:deflong-replica-edm}
\end{equation}
Thus the underlying manifold of $\mathcal J_n(\varnothing)$ contains $n$ copies $J_{k,+}(\sigma)$ with the original data and $n$ copies $J_{k,-}(\sigma)$ with the involuted data, where $k=1,\ldots,n$.
It also contains $2n$ gluing joints, one at each cutoff circle.

In the notation of Sec.~\ref{sec:reduce}, the corresponding replica trace is
\begin{equation}
 \tr_\chi[\tilde\rho(\chi,\bar\chi)^n]
 =G[\mathcal J_n(\varnothing)]
 \stackrel{G_N\to0}{\approx}e^{-I_n},
 \label{eq:deflong-replica-gpi}
\end{equation}
where $I_n$ is the Euclidean on-shell action of a saddle filling the closed EDM $\mathcal J_n(\varnothing)$.
When computing the entropy, we also need to divide this trace by the normalization of the density matrix: the division by $[\tr_\chi\tilde\rho(\chi,\bar\chi)]^n$ is implemented below by subtracting $nI_1$.

\subsection{Solving the Boundary Value Problem}
\label{subsec:deflong-embedding}
To evaluate the replica trace, we seek an on-shell geometry containing a slice $J(\sigma)$ with the prescribed elliptic data, which consist of the trace $K(x)$ and the conformal class.
Locally, an embedded surface is fully characterized by its intrinsic metric and extrinsic curvature.
Our boundary condition fixes only the conformal class of the induced metric $h_{ij}$ and the trace $K$ of the extrinsic curvature $K_{ij}$, and the boundary value problem aims to specify the remaining components to fully determine its embedding in an on-shell bulk geometry.
To do this, we first use the momentum and Hamiltonian constraints, which are differential equations on the phase space data, to determine the remaining intrinsic and extrinsic data locally, leaving some integration constants.
We then fix these constants by matching the prescribed conformal modulus and requiring the $2n$ surfaces $J_{k,\pm}(\sigma)$ to close smoothly around the Euclidean-time circle.

\paragraph{Local embedding of one $J(\sigma)$.}
In three-dimensional vacuum gravity the filling is locally AdS$_3$, and rotational symmetry reduces the constraints on $J_{\pm}(\sigma)$ to radial ordinary differential equations.
We first use the circumference radius $r$ as the radial coordinate, so that
\begin{equation}
 ds_{\Sigma_n}^2=h_{rr}^{(n)}(r)dr^2+r^2d\phi^2,
 \qquad
 K^i{}_{j,n}=\operatorname{diag}(K^r{}_{r,n},K^\phi{}_{\phi,n}).
 \label{eq:deflong-areal-data}
\end{equation}
The conformal boundary condition holds the trace fixed as a function of a marked conformal coordinate.
Explicitly, we bring the induced metric to conformal gauge by writing $r=\varrho_n(x)$ and choosing $x$ such that
\begin{equation}
 \frac{dx}{dr}=\frac{\sqrt{h_{rr}^{(n)}(r)}}{r}.
 \label{eq:deflong-conformal-metric}
\end{equation}
It follows that $ds_{\Sigma_n}^2=\varrho_n(x)^2(dx^2+d\phi^2)$, so $dx^2+d\phi^2$ is the chosen representative of the conformal class, and the trace of the extrinsic curvature must equal the prescribed function $K(x)$.
We keep $n$ explicit in $r=\varrho_n(x)$ to emphasize that the relation between the physical circumference $r$ and the conformal marking coordinate $x$ depends on $n$.
The intrinsic and extrinsic data in this gauge therefore contain three functions: $\varrho_n(x)$ and the two diagonal components of $K^i{}_{j,n}$.
The trace condition,
$K(x)=K^x{}_{x,n}(x)+K^\phi{}_{\phi,n}(x)$, together with the momentum and Hamiltonian constraints, gives three equations that fix these three unknown functions locally up to integration constants.

The trace and momentum constraints first determine the two components of the extrinsic curvature as functions of $\varrho_n$.
The momentum constraint is
\begin{equation}
D_j\left(K^j{}_{i,n}-\delta^j_i K\right)=0
\quad\Longrightarrow\quad
\frac{d}{dx}\left(\varrho_n^2K^\phi{}_{\phi,n}\right)
=\varrho_n\varrho_n'K(x).
\end{equation}
We define the first integral of this equation by
\begin{equation}
\begin{aligned}
\mathcal P_n(x)\equiv\varrho_n^2K^\phi{}_{\phi,n}
&=\mathcal E_n
+\int_0^x d\widetilde x\,
\varrho_n(\widetilde x)\varrho_n'(\widetilde x)K(\widetilde x).
\end{aligned}
\label{eq:deflong-K-components}
\end{equation}
It follows that
$K^\phi{}_{\phi,n}=\mathcal P_n/\varrho_n^2$ and
$K^x{}_{x,n}=K(x)-\mathcal P_n/\varrho_n^2$.
Here a prime denotes an $x$ derivative, and
$\mathcal E_n=\mathcal P_n(0)$ is a generally complex integration constant.

The remaining function $\varrho_n$ is fixed by the Hamiltonian constraint.
Using \eqref{eq:deflong-K-components}, the constraint becomes
\begin{equation*}
R[h]-K^2+K^i{}_{j,n}K^j{}_{i,n}+2=0
\quad\Longrightarrow\quad
\left(\frac{\varrho_n'}{\varrho_n} \right)'
=\varrho_n^2-\mathcal P_nK(x)
+\frac{\mathcal P_n^2}{\varrho_n^2}.
\end{equation*}
Combining this equation with the momentum constraint and integrating once, we obtain
\begin{equation}
h_{rr}^{(n)}\big|_{r=\varrho_n(x)}
=\frac{\varrho_n^2}{(\varrho_n')^2}
=\frac{\varrho_n^2}
{\varrho_n^4-r_+^2\varrho_n^2-\mathcal P_n^2},
\label{eq:deflong-areal-family-main}
\end{equation}
where $r_+^2$ is the second integration constant.
Choosing the square-root branch continued from the norm saddle and eliminating $\mathcal P_n$ from \eqref{eq:deflong-areal-family-main} gives a single second-order equation for $\varrho_n$:
\begin{equation}
\varrho_n''
=
2\varrho_n^3-r_+^2\varrho_n
-\varrho_n K(x)
\sqrt{
\varrho_n^4-r_+^2\varrho_n^2-(\varrho_n')^2
}.
\label{eq:deflong-varrho-odes}
\end{equation}
When the prescribed trace is constant, the constraints simplify further.
At the minimal-area circle $x=0$, we impose $\varrho_n'(0)=0$ and $\varrho_n(0)=\varrho_{*,n}$.\footnote{
If $K(x)=K$, the momentum constraint can be integrated explicitly:
\begin{align*}
 \mathcal P_n(x)
 &=\mathcal E_n+\frac{K}{2}
 \left[\varrho_n(x)^2-\varrho_{*,n}^2\right],\\
 (\varrho_n')^2
 &=\varrho_n^4-r_+^2\varrho_n^2
 -\left\{
 \mathcal E_n+\frac{K}{2}
 \left[\varrho_n^2-\varrho_{*,n}^2\right]
 \right\}^2.
\end{align*}
The local constraints therefore reduce to a single separable first-order equation, again with $\mathcal E_n^2=\varrho_{*,n}^4-r_+^2\varrho_{*,n}^2$; no separate Liouville-type equation is required.
For $K=0$, this reduces to the static system studied in the companion Letter~\cite{shortpaper}.}
For fixed $(r_+,\mathcal E_n)$, the turning relation $\mathcal E_n^2=\varrho_{*,n}^4-r_+^2\varrho_{*,n}^2$ determines $\varrho_{*,n}$ on the branch continued from the $n=1$ norm saddle, and these initial conditions determine the local solution of \eqref{eq:deflong-varrho-odes}.

To completely determine the embedding of a rotationally symmetric slice, one needs to specify $r=\varrho_n(x)$ and $\tau=\tau_n(x)$.
Given $r=\varrho_n(x)$, we find
\begin{equation}
 \tau_n'(x)=\frac{\mathcal P_n(x)}{\varrho_n(x)^2-r_+^2}
 \label{eq:deflong-Tprime}
\end{equation}
to obtain the induced metric $ds_{\Sigma_n}^2=\varrho_n(x)^2(dx^2+d\phi^2)$.
We have therefore obtained a family of embeddings parametrized by $r_+$ and $\mathcal{E}_n$.

\paragraph{Global matching and closure of $n$ replicas. }
The local solution still depends on the two parameters $r_+$ and $\mathcal E_n$.
We now fix them by matching the prescribed conformal moduli and requiring the replica geometry to close around the Euclidean-time circle.

The marked conformal coordinate spans $-\pi m_R\leq x\leq\pi m_R$.
${\varrho}_n$ is parametrized by $r_+$ and  $\mathcal E_n$, and the first condition is that 
\begin{equation}
 \varrho_n(\pi m_R)=R.
 \label{eq:deflong-modmatch}
\end{equation}
For each trial value of $r_+$, this condition fixes $\mathcal E_n$, and hence $\varrho_{*,n}$ through the turning-point relation, on the branch continued from the norm saddle.

After this matching, only $r_+$ remains to be fixed.
The embedding equation gives a generally complex Euclidean-time displacement between the two ends of the $J_+(\sigma)$ slice, while the involuted slice $J_-(\sigma)$ gives its complex conjugate.
On the branch continued from the norm saddle, the corresponding positive real increment is
\begin{equation}
 \Delta\tau(r_+,\mathcal E_n)
 =2\operatorname{Re}\!
 \int_0^{\pi m_R}dx\,
 \frac{\mathcal P_n(x)}{\varrho_n(x)^2-r_+^2}.
 \label{eq:deflong-Deltatau}
\end{equation}
The closed EDM $\mathcal J_n(\varnothing)$ contains $n$ original--involuted pairs.
Each pair contributes $2\Delta\tau$ to the Euclidean-time circle, so the surfaces form a smooth BTZ filling only when their total increment equals the thermal period $2\pi/r_+$.
This gives the final condition
\begin{equation}
 2n\Delta\tau=\frac{2\pi}{r_+}.
 \label{eq:deflong-closure}
\end{equation}
Solving this condition fixes $r_+(n)$; the endpoint condition then gives $\mathcal E_n$ and $\varrho_{*,n}$.
At $n=1$, these equations reproduce the original norm saddle with $r_+=r_1$.
In the static limit $\mathcal E_0,\epsilon_R,K\to0$, the selected integration constant $\mathcal E_n$ reduces to the real quantity $E_n$ used in the companion Letter~\cite{shortpaper}.

\paragraph{Thermal AdS.}
We also consider a filling with thermal-AdS topology.
The same constraint and embedding equations apply after setting
\begin{equation}
 r_+^2=-1.
 \label{eq:deflong-ads-mass}
\end{equation}
We denote the corresponding functions by $\varrho_{\rm AdS}(x)$ and $\mathcal P_{\rm AdS}(x)$.
The modulus condition fixes $\varrho_{*,\rm AdS}$ and hence $\mathcal E_{\rm AdS}$, and the embedding integral determines $\Delta\tau_{\rm AdS}$.
Since the $\phi$-circle contracts at the global-AdS origin while Euclidean time remains noncontractible, there is no horizon or BTZ closure condition.
The replica geometry consequently has Euclidean period $\beta_{\rm AdS}(n)=2n\Delta\tau_{\rm AdS}$, and the action computed below is linear in $n$.

\subsection{On-Shell Action and Entropy}
\label{subsec:deflong-action}

\paragraph{Bulk and boundary terms.}
We first evaluate the Einstein--Hilbert term in the bulk and the conformal-boundary terms on the preparation manifolds $J_{\pm}(\sigma)$; the Hayward terms at the cutoff joints are treated below.
After dividing the replica-symmetric action by $n$, the quotient is bounded by $J_+(\sigma)$, carrying the original data, and $J_-(\sigma)$, carrying the involuted data.
For $\varrho_{*,n}\leq r\leq R$, their Euclidean-time separation is determined by
\begin{equation*}
 \Delta\tau\bigl(\varrho_n(x)\bigr)
 =2\operatorname{Re}\int_x^{\pi m_R}d\widetilde x\,
 \frac{\mathcal P_n(\widetilde x)}
 {\varrho_n(\widetilde x)^2-r_+^2},
\end{equation*}
and we extend $\Delta\tau(r)$ as a constant for $r_+\leq r\leq\varrho_{*,n}$.
Since the conformal-boundary term in three bulk dimensions has one half of the usual Gibbons--Hawking coefficient, the on-shell relation $R[g]-2\Lambda=-4$ gives
\begin{align}
 -\frac{8\pi G_N}{n}(I_{\rm EH}+I_{\rm CBC})
 &=-8\pi\int_{r_+}^{R}dr\,r\Delta\tau(r)
 \notag\\
 &\quad+\pi\sum_{\eta=\pm}\int_{J_\eta(\sigma)}dr\,r\sqrt{h_{rr}}\,K.
 \label{eq:deflong-action-first-step}
\end{align}
Because $r$ runs from $R$ to $\varrho_{*,n}$ and back to $R$ along each $J_\eta(\sigma)$, the last term is the sum of two integrals over $\varrho_{*,n}\leq r\leq R$.
On these two sides, the embedding equation \eqref{eq:deflong-Tprime} gives $r\sqrt{h_{rr}}K^\phi{}_{\phi}=\pm(r^2-r_+^2)d\tau/dr$, with the sign fixed by the outward orientation.
After summing both sides of $J_+(\sigma)$ and $J_-(\sigma)$, their contribution is $-\int_{\varrho_{*,n}}^{R}dr\,(r^2-r_+^2)d\Delta\tau/dr$.
We then split $K=(K^r{}_r-K^\phi{}_{\phi})+2K^\phi{}_{\phi}$ and integrate the bulk term by parts, giving
\begin{align}
 -\frac{8\pi G_N}{n}(I_{\rm EH}+I_{\rm CBC})
 &=-4\pi\int_{r_+}^{R}d\!\left(r^2\Delta\tau(r)\right)
 +4\pi r_+^2\int_{\varrho_{*,n}}^{R}dr\,\frac{d\Delta\tau}{dr}
 \notag\\
 &\quad+\pi\sum_{\eta=\pm}\int_{J_\eta(\sigma)}dr\,
 r\sqrt{h_{rr}}\left(K^r{}_r-K^\phi{}_{\phi}\right).
 \label{eq:deflong-action-parts}
\end{align}
The two terms in the first line cancel: $\Delta\tau(R)=0$, while $\Delta\tau(r)$ is constant below $\varrho_{*,n}$, so they are respectively $4\pi r_+^2\Delta\tau(\varrho_{*,n})$ and $-4\pi r_+^2\Delta\tau(\varrho_{*,n})$.
We are therefore left with the last integral in \eqref{eq:deflong-action-parts}.
Writing the integral on the right side of $J_+(\sigma)$ in terms of $r=\varrho_n(x)$, Eqs.~\eqref{eq:deflong-K-components} and \eqref{eq:deflong-areal-family-main} give
\begin{equation*}
 dr\,r\sqrt{h_{rr}^{(n)}}
 \left(K^r{}_{r,n}-K^\phi{}_{\phi,n}\right)
 =dx\left[\varrho_n(x)^2K(x)-2\mathcal P_n(x)\right].
\end{equation*}
In the left--right-symmetric setup, the two sides of $J_+(\sigma)$ contribute equally, while the two sides of $J_-(\sigma)$ give their complex conjugates.
Restoring the $n$ pairs in the replica cover gives
\begin{equation}
 G_N\bigl(I_{\rm EH}+I_{\rm CBC}\bigr)_n^{\rm BTZ}
 =-\frac{n}{2}\operatorname{Re}
 \int_0^{\pi m_R}dx\,
 \left[\varrho_n(x)^2K(x)-2\mathcal P_n(x)\right].
 \label{eq:deflong-action-reduced-n}
\end{equation}
For thermal AdS, the derivation differs only at the inner endpoint.
Because the $\phi$-circle, rather than the Euclidean-time circle, contracts at the origin, the integration by parts in \eqref{eq:deflong-action-parts} gives the additional endpoint term shown below:\footnote{As a check, when $K=0$, $\mathcal{P}$ is a (real) constant. 
We still use $\mathcal{E}$ to denote these constants: For the BTZ embedding of the $n$-th replica, $\mathcal P_n=\mathcal{E}_n$, and for the AdS embedding $\mathcal P_{\rm AdS}=\mathcal E_{\rm AdS}$, and these expressions reduce to
\[
 G_N\bigl(I_{\rm EH}+I_{\rm CBC}\bigr)_n^{\rm BTZ}
 =n\pi m_R \mathcal{E}_n,
 \qquad
 G_N\bigl(I_{\rm EH}+I_{\rm CBC}\bigr)_n^{\rm AdS}
 =n\left(\pi m_R\mathcal E_{\rm AdS}-\frac{\Delta\tau_{\rm AdS}}{2}\right),
\]
which are the actions used in the companion Letter~\cite{shortpaper}.}
\begin{equation}
\begin{aligned}
 G_N\bigl(I_{\rm EH}+I_{\rm CBC}\bigr)_n^{\rm AdS}
 &=-\frac{n}{2}\operatorname{Re}\Biggl[
 \int_0^{\pi m_R}dx\,
 \left[\varrho_{\rm AdS}(x)^2K(x)-2\mathcal P_{\rm AdS}(x)\right]
 \\[-2pt]
 &\qquad{}+\Delta\tau_{\rm AdS}\Biggr].
\end{aligned}
 \label{eq:deflong-ads-origin-correction}
\end{equation}

\paragraph{Hayward terms at finite cutoff.}
At finite $R$, $J_+(\sigma)$ meets $J_{-}(\sigma)$ at each of the cutoff circles $\chi$ and $\chi_c$.
The action therefore includes a Hayward term at each circle~\cite{Hayward:1993my}.
We use the convention of \eqref{eq-IEinstein}: if $\alpha_\chi$ and $\alpha_{\chi_c}$ are the interior angles measured through the filled bulk region, the signed exterior angles are
\begin{equation}
\begin{alignedat}{2}
 \Theta_\chi&=\pi-\alpha_\chi,
 &\qquad \Theta_{\chi_c}&=\pi-\alpha_{\chi_c},\\
 I_{n,H}&=-n\frac{2\pi R}{8\pi G_N}
 \left[\Theta_\chi(n)+\Theta_{\chi_c}(n)\right].
\end{alignedat}
 \label{eq:deflong-Hayward-oriented}
\end{equation}
The factor of $n$ counts the $n$ copies of each cutoff circle.
The filled side fixes the sign: $\Theta=0$ for a smooth gluing, while $\Theta$ is positive for a convex corner and negative for a concave corner.

To evaluate either angle, consider the rotationally symmetric metric
\begin{equation*}
ds^2=f(r)d\tau^2+\frac{dr^2}{f(r)}+r^2d\phi^2,
\end{equation*}
and write the two meeting slices as $\tau=\tau_\pm(r)$.
In this paragraph, primes on $\tau_\pm$ denote derivatives with respect to $r$.
Their induced radial metrics and outward unit normal covectors are
\begin{equation}
 h_{rr}^{\pm}=f^{-1}+f(\tau_\pm')^2,
 \qquad
 n_\mu^{\pm}=\pm\frac{(1,-\tau_\pm',0)}{\sqrt{h_{rr}^{\pm}}},
 \label{eq:deflong-joint-normal}
\end{equation}
where the signs are fixed by requiring both normals to point out of the filled region.
It follows that
\begin{equation}
 \cos\Theta=n^+\!\cdot n^-
 =-
 \frac{f^{-1}+f\tau_+'\tau_-'}
 {\sqrt{h_{rr}^{+}h_{rr}^{-}}}.
 \label{eq:deflong-joint-normal-product}
\end{equation}
For a real saddle, the scalar product determines $|\Theta|$, while the filled side determines its sign.
For the complex replica saddles, the corresponding branch of the inverse cosine is fixed by analytic continuation from $n=1$.


For the conjugate gluing used here, both exterior angles approach $\pi$ as $R\to\infty$, so each unnormalized Hayward action contains a divergent term proportional to $R$.
This common term cancels in the normalized combination $I_{n,H}-nI_{1,H}$.
For the cutoff profile used here,\footnote{Here $\tau_\pm'(R)=O(R^{-2})$ and $n^+\!\cdot {n}^-=-1+O(R^{-4})$, so the difference between a replica angle and its $n=1$ value is $O(R^{-2})$.
Multiplication by the circumference $2\pi R$ then gives an $O(R^{-1})$ action.
This scaling depends on the profile and boundary condition.}
\begin{equation}
 I_{n,H}-nI_{1,H}=O(R^{-1}).
 \label{eq:deflong-corner-suppression}
\end{equation}
Hence the normalized Hayward contribution vanishes when the cutoff is removed.

\begin{figure}[tbp]
    \begin{minipage}{0.95\linewidth}
        \centering
        \includegraphics[width=0.98\linewidth]
        {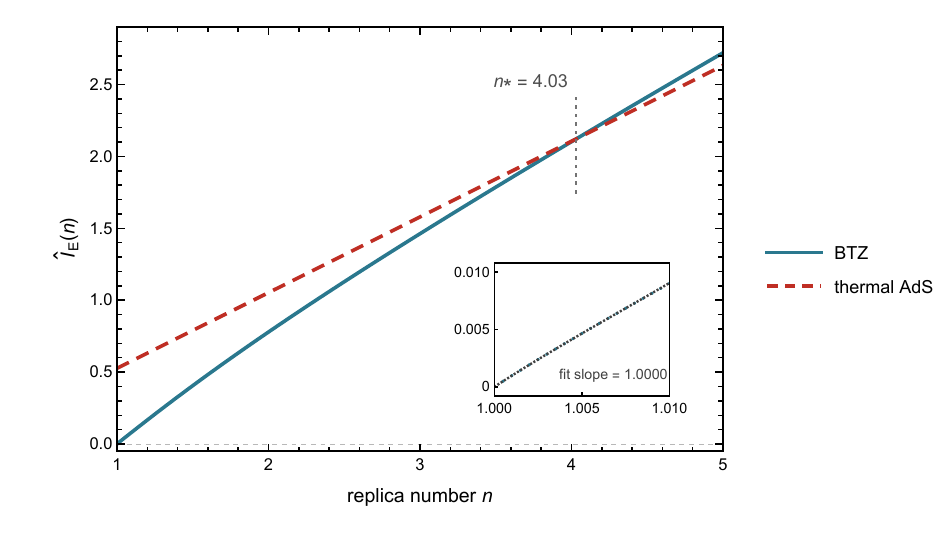}
    \begin{figurecaptionblock}
        \caption[Normalized Euclidean BTZ and thermal-AdS actions]{Normalized Euclidean BTZ and thermal-AdS actions defined in \eqref{eq:deflong-action-normalized}, with $(r_a,r_1,R,\epsilon)=(2,\sqrt{5},500,10^{-2})$.
        BTZ dominates near $n=1$, while thermal AdS dominates beyond the crossing at $n_*=4.03$ marked by the vertical dashed line.
        The inset resolves the BTZ branch over $1\le n\le 1.01$; the points are the computed actions, and the dotted fit has linear coefficient $c_1=1.0000$, reproducing the normalized von Neumann entropy in \eqref{eq:deflong-entropy-target}.}
        \label{fig:deflong-action-vs-n}
    \end{figurecaptionblock}
    \end{minipage}
\end{figure}

\begin{figure}[tbp]
    \begin{minipage}{0.95\linewidth}
        \centering
        \includegraphics[width=0.98\linewidth]{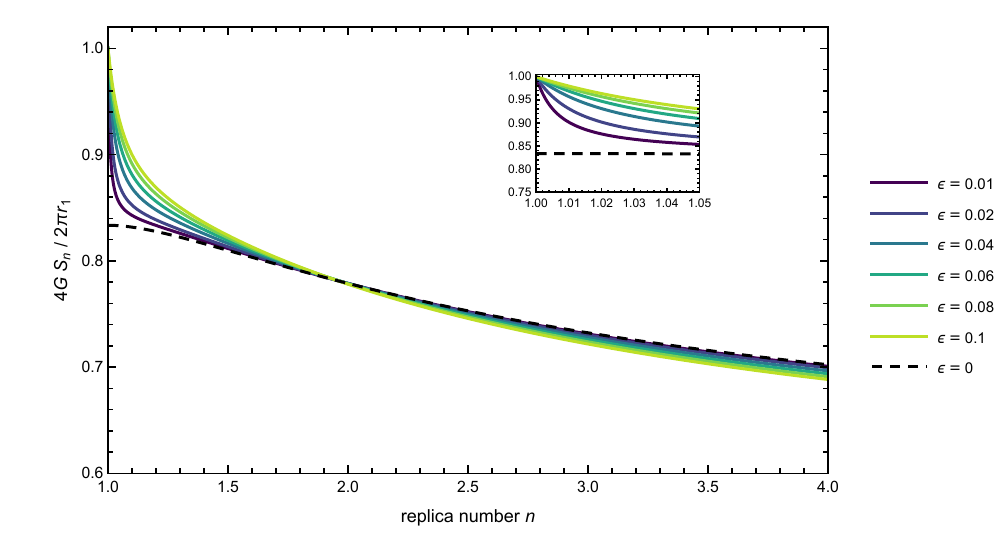}
    \begin{figurecaptionblock}
        \caption[Normalized R\'enyi entropy]{ %
            The R\'enyi entropy $S_n\bigl[\rho(\chi,\bar\chi)\bigr]$ has the universal limit $S_1=2\pi r_1/(4G_N)$ as $n\to 1$, independently of the amplitude $\epsilon$ of the complex deformation~\eqref{eq:deflong-profile-graph}.
            With  $\epsilon=0$, however, the $n\to 1$ limit yields instead the entropy of the BTZ black hole containing a time-symmetric slice with the undeformed data $(r_a,r_1,R)=(2,\sqrt{5},500)$.
            This illustrates the importance of a nonzero complex deformation.
        }
        \label{fig:deflong-renyi-vs-n}
    \end{figurecaptionblock}
    \end{minipage}
\end{figure}

\paragraph{Saddle comparison and entropy.}
Let $I_n^{\rm BTZ}$ and $I_n^{\rm AdS}$ denote the total Euclidean actions obtained by adding the Hayward terms to \eqref{eq:deflong-action-reduced-n} and \eqref{eq:deflong-ads-origin-correction}.
Equation~\eqref{eq:deflong-corner-suppression} shows that the normalized Hayward contributions vanish at large $R$, so the large-cutoff comparison is determined by the two smooth-boundary actions.
We normalize the replica trace by subtracting $n$ copies of the $n=1$ BTZ action $I_1$.\footnote{Here $I_1$ denotes the BTZ saddle, which gives the leading contribution to $Z_1=e^{-I_1}$ at the semiclassical order considered here.
A thermal-AdS saddle also exists at $n=1$, but it is not a small complex deformation of the Lorentzian preparation slice.}
\begin{align}
 \widehat I^{\rm BTZ}(n)
 &=\frac{4G_N}{2\pi r_1}\left[I_n^{\rm BTZ}-nI_1\right],
 \notag\\
 \widehat I^{\rm AdS}(n)
 &=\frac{4G_N}{2\pi r_1}\left[I_n^{\rm AdS}-nI_1\right].
 \label{eq:deflong-action-normalized}
\end{align}
At leading order, the saddle with the smaller normalized action dominates, and the normalized R\'enyi entropy is\footnote{This comparison assumes that both saddles contribute on the chosen contour and that their one-loop prefactors do not reverse the classical ordering away from the crossing.}
\begin{equation}
 \frac{4G_N}{2\pi r_1}S_n
 =\frac{1}{n-1}\min\left\{
 \widehat I^{\rm BTZ}(n),\widehat I^{\rm AdS}(n)\right\}.
 \label{eq:deflong-renyi}
\end{equation}
Figure~\ref{fig:deflong-action-vs-n} shows this comparison directly: the BTZ saddle that reduces to the norm geometry at $n=1$ dominates near $n=1$, while thermal AdS takes over at the crossing of the two actions.
The inset resolves the BTZ action close to $n=1$ and verifies
\begin{equation}
 \frac{d\widehat I^{\rm BTZ}}{dn}\bigg|_{n=1}=1,
 \qquad
 \lim_{n \to 1}S_n=\frac{2\pi r_1}{4G_N}.
 \label{eq:deflong-entropy-target}
\end{equation}
Thus the von Neumann entropy is set by the horizon circumference $2\pi r_1$ of the original norm geometry, rather than by the circumference $2\pi r_a$ of the smallest circle on the $\Sigma$ slice (the original undeformed Lorentzian partial Cauchy slice).
Figure~\ref{fig:deflong-renyi-vs-n} gives a complementary check by plotting the BTZ R\'enyi entropy for several values of $\epsilon$.
The plotted BTZ branch dominates near $n=1$; at larger $n$, the thermal-AdS saddle must be included through the minimum in \eqref{eq:deflong-renyi}.
The R\'enyi entropies depend on the deformation away from $n=1$, but they all approach the same value in the $n\to 1$ limit.
The von Neumann entropy is therefore independent of (nonzero) $\epsilon$, even though the deformation affects the higher R\'enyi entropies and the replica geometry.

\begin{figure}[tbp]
    \begin{minipage}{0.95\linewidth}
        \centering
        \includegraphics[width=0.98\linewidth]
        {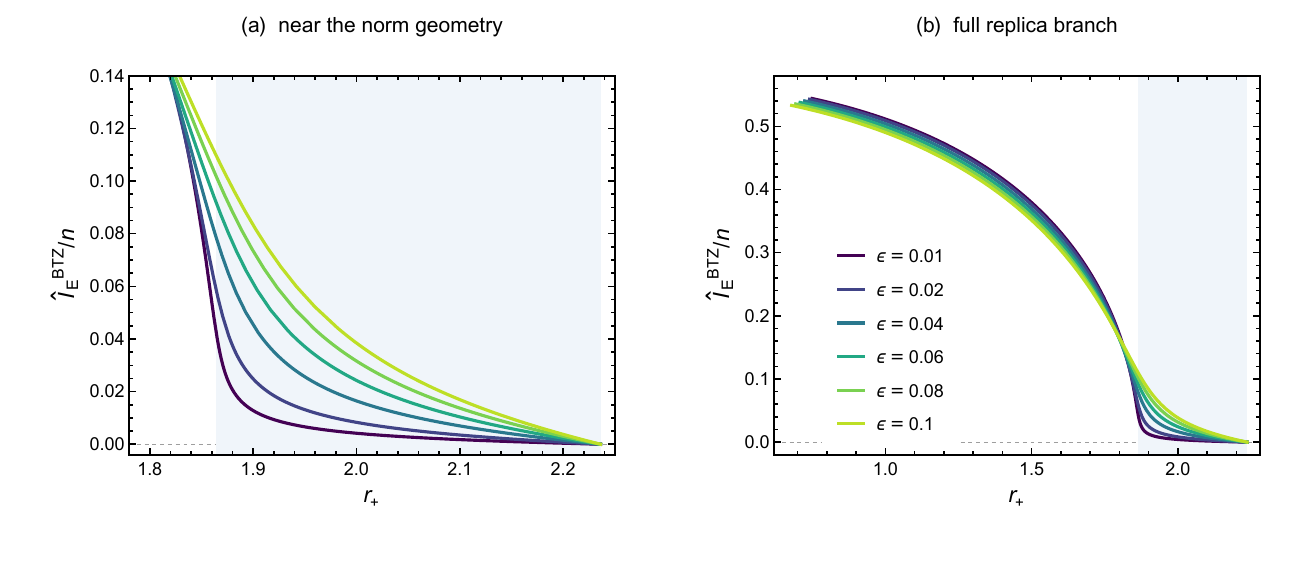}
    \begin{figurecaptionblock}
        \caption[Normalized quotient action]{The normalized quotient action $\widehat I^{\rm BTZ}(n)/n$ is plotted against the BTZ radius $r_+$ for the same six deformation amplitudes as in Fig.~\ref{fig:deflong-renyi-vs-n}.
        The light-blue region is the radius interval between the minimal radius of the modulus-matched $K=0$ slice and $r_1$.\protect\footnotemark
        Across this interval the action changes only weakly while $r_+$ moves by an order-one amount, which is the numerical remnant of the Lorentzian flat direction.
        Near the left edge the curves turn onto the branch with a finite Euclidean region and develop an appreciable slope.
        Panel (a) resolves the nearly flat region, while panel (b) shows the full replica branch.}
        \label{fig:deflong-action-vs-rplus}
    \end{figurecaptionblock}
    \end{minipage}
\end{figure}

\paragraph{Deformation dependence and geometric interpretation.}
We now explain the $\epsilon$ dependence in Fig.~\ref{fig:deflong-renyi-vs-n} using the closure condition.

The common limit at $n=1$ and the rapid separation of the curves away from it both follow from how the closure condition \eqref{eq:deflong-closure} fixes $r_+$.
After imposing modulus matching, we may regard $\Delta\tau$ as a function of $r_+$ and $\epsilon$; differentiating $r_+\Delta\tau/\pi=1/n$ with respect to $n$ at fixed $\epsilon$ gives
\begin{equation}
 \frac{\partial r_+}{\partial n}
 =-\frac{1}{n^2\,\partial_{r_+}(r_+\Delta\tau/\pi)}.
 \label{eq:deflong-closure-derivative}
\end{equation}
At $\epsilon=0$, the Euclidean separation of every horizon-crossing embedding comes entirely from the prescribed bypass of the horizon pole, so $r_+\Delta\tau/\pi=1$ is independent of $r_+$.
The $n=1$ closure condition therefore leaves $r_+$ undetermined along this Lorentzian family, and the denominator in \eqref{eq:deflong-closure-derivative} vanishes.
This is a flat direction of the undeformed closure problem.
A nonzero mixed-conformal deformation lifts it at order $\epsilon$: the denominator becomes $O(\epsilon)$ and $\partial_n r_+=O(\epsilon^{-1})$.
Thus, within the narrow range $n-1=O(\epsilon)$, $r_+$ changes by an order-one amount while the saddle still approaches $r_+=r_1$ in the von Neumann limit.
This rapid motion separates the R\'enyi curves near $n=1$ in Fig.~\ref{fig:deflong-renyi-vs-n}, while the action remains nearly flat over the corresponding range of $r_+$ in Fig.~\ref{fig:deflong-action-vs-rplus}.

\footnotetext{The left endpoint is fixed by matching the undeformed cylinder modulus to the real $K=0$ outer branch: $\frac{1}{\pi r_{\rm out}}\arccos\frac{r_{\rm out}}{R}=\frac{1}{\pi}\int_{r_a}^{R} dr/\sqrt{(r^2-r_a^2)[r^2-(r_1^2-r_a^2)]}$.
For $(r_a,r_1,R)=(2,\sqrt{5},500)$, this gives $r_{\rm out}=1.863$, which is shown approximately in the figure.}

Outside this range, where $n-1\gg\epsilon$, the $O(n-1)$ change in the right-hand side of the closure condition cannot be produced by the $O(\epsilon)$ correction to the Lorentzian family.
The saddle instead approaches the real $K=0$ BTZ branch, for which $r_+<r_a$ and the $J(\sigma)$ slice remains outside the horizon.
At $\epsilon=0$, this branch already has a finite Euclidean-time extent, so the deformation gives only a small complex correction.
Near its endpoint, the undeformed change in $r_+$ is $O((n-1)^2)$, while the deformation contributes $O(\epsilon(n-1))$; its relative effect is therefore $O(\epsilon/(n-1))$.
Although $\partial_n r_+$ is large within the range $n-1=O(\epsilon)$, the action remains regular because the implicit $n$-dependence of the saddle fields drops out of its first variation.
The replica relation identifies the explicit replica derivative of the quotient action with the finite circumference $2\pi r_+$~\cite{Lewkowycz:2013nqa,Dong:2016fnf}.
Thus an order-one change in $r_+$ over an interval of width $O(\epsilon)$ changes the action only by $O(\epsilon)$.
A nonzero $\epsilon$ therefore selects a smooth continuation of the BTZ saddle to the norm geometry without changing the von Neumann entropy.
For $n-1\gg\epsilon$, its effect is small, and the later BTZ--thermal-AdS crossing is a separate competition between topologies.

\appendix

\section{Resolution of an Apparent Counterexample to Positivity}
\label{app:positivity_examples}

\begin{figure}[tbp]
    \centering
   \includegraphics[width=0.5\textwidth]{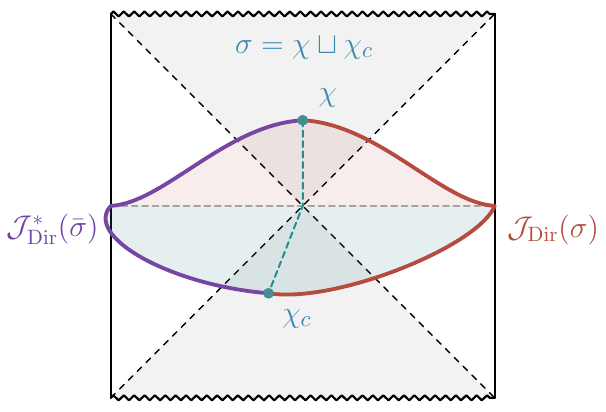}
    \begin{figurecaptionblock}
    \caption[The Wall geometry]{Complex geometry consisting of a real Euclidean section (blue) joined to a real Lorentzian section (orange). Dirichlet data manifolds $\j_{\rm Dir}(\sigma)$ and $\j_{\rm Dir}^*(\bar\sigma)$ are shown in red and purple, respectively. This solution and its CRT conjugate dominate the GPI contribution to the norm of $\j_{\rm Dir}(\sigma)$; this makes the GPI oscillatory, leading to negative-norm states~\cite{Wall:2021bxi}. If the Dirichlet data are replaced by the mixed-conformal data induced by the same saddles, the same complex saddle pair still contributes to the GPI. But for generic parameters this GPI no longer represents a norm, and in the special case when it does, a new real Euclidean saddle dominates (see Fig.~\ref{fig:wall-fully-euclidean-rewrite}) and restores positivity.}
    \label{fig:wall-norm-schematic}
    \end{figurecaptionblock}
\end{figure}

As noted in Sec.~\ref{sec:outlook}, our proposal rests on the absence of negative norm states, at least with \emph{some} choice of elliptic data type. With Dirichlet data, Wall~\cite{Wall:2021bxi} was able to construct a negative-norm state $\j_{\rm Dir}(\sigma)$ on a cylinder (see Fig.~\ref{fig:wall-norm-schematic}). The norm-GPI is dominated not by a CRT-invariant saddle (which would have real action), but by a \emph{pair} of saddles related by CRT, with actions $I$ and $\overline I$:
\begin{equation}
\begin{aligned}
G[(\mathcal J^*_{\rm Dir})_A(\bar\sigma) \mathcal J_{\rm Dir}^A(\sigma)]
&=e^{-I}+e^{-\overline I}\\
&=2e^{-\operatorname{Re}I}\cos\!\left(\operatorname{Im}I\right).
\end{aligned}
\label{eq:general-CRT-pair}
\end{equation}
For nonzero $\operatorname{Im}(I) \sim 1/G_N$, the cosine oscillates rapidly and can be negative.

A CRT-breaking pair of saddles would pose no threat to positivity if they were subleading to a positive contribution. However, no additional saddles are known that would dominate over the pair of Wall saddles and restore positivity for the norm of $\j_{\rm Dir}(\sigma)$. This is a key reason why we reject Dirichlet data\footnote{This is not the interpretation taken in Ref.~\cite{Wall:2021bxi}.} and formulate our proposal in terms of mixed-conformal data instead. (A second reason, unrelated to this Appendix, is that Dirichlet data are not, in fact, elliptic in general~\cite{Witten:2018lgb}.)

It is not obvious, of course, that positivity holds when states are defined as mixed-conformal elliptic data manifolds. Here we will examine a narrower question: can a negative-norm state be constructed from mixed-conformal boundary data \emph{induced by the Wall saddles}?

Generically, we will find that the answer is no: the mixed-conformal boundary conditions induced by the Wall saddles do not correspond to a norm, just to an inner product that happens to be negative.
And in a non-generic case where such an interpretation exists, we will find that the answer is still no, because in this case the norm-GPI admits another saddle with smaller real part of the action that restores positivity.



\paragraph{Dirichlet elliptic data manifolds with negative norm}

We begin by reviewing the Wall state $\j_{\rm Dir}(\sigma)$~\cite{Wall:2021bxi}, using our terminology and notational conventions. For concreteness, we specialize Wall's AdS--Schwarzschild construction to the nonrotating BTZ solution ($d=2$). The state consists of Dirichlet data induced on a submanifold of its complexification. The Euclidean metric is again
\begin{equation}
ds^2 = f(r)\, d\tau^2 + \frac{dr^2}{f(r)} + r^2 d\phi^2,
\qquad
f(r)=r^2-r_h^2,
\end{equation}
with $\tau\sim \tau+\beta$, and $\beta=2\pi/r_h$.
Analytic continuation from the time-reflection-symmetric slice gives the Lorentzian section
\begin{equation}
ds^2 = -f(r)\,dt^2 + \frac{dr^2}{f(r)} + r^2d\phi^2.
\end{equation}

Now consider a closed submanifold $K_{\rm Dir}(\varnothing)$ of this complexified geometry (see Fig.~\ref{fig:wall-norm-schematic}), consisting of the cylinder $0\leq \tau\leq \beta/2$, $r=R_c$, in the Euclidean section, joined at $t=0$ to a smooth, left--right-symmetric Cauchy slice in the Lorentzian section that lies strictly to the future of the time-reflection symmetric slice and thus passes through the black hole interior.

By cutting this submanifold into two halves exchanged by left--right reflection, we obtain manifolds $J_{\rm Dir}(\sigma)$ and $J^*_{\rm Dir}(\bar\sigma)$. Here $\sigma$ is the union of the circle $r=R_c, \tau=\beta/4$ in the Euclidean sector and the circle at the midpoint of the Lorentzian Cauchy slice. Equipping $J_{\rm Dir}(\sigma)$ with its induced Dirichlet data defines the quantum state $\mathcal J^A_{\rm Dir}(\sigma)$. Similarly, the oppositely oriented half $J^*_{\rm Dir}(\bar\sigma)$ may be equipped with its induced Dirichlet data. The Dirichlet data on both halves are manifestly identical and real, so CRT exchanges the two halves and the induced data on $J^*_{\rm Dir}(\bar\sigma)$ define the dual state $(\mathcal J^*_{\rm Dir})_A(\bar\sigma)$.

The closed manifold $K_{\rm Dir}(\varnothing)$ is obtained by gluing $J_{\rm Dir}(\sigma)$ to $J^*_{\rm Dir}(\bar\sigma)$ along $\sigma$.
The norm of $\mathcal J^A_{\rm Dir}(\sigma)$ is computed by evaluating the GPI $G[(\mathcal J^*_{\rm Dir})_A(\bar\sigma) \mathcal J_{\rm Dir}^A(\sigma)]$.
The GPI can be approximated semiclassically by the saddle-point method.

There are two obvious saddles. The geometry described above and shown in Fig.~\ref{fig:wall-norm-schematic} is a solution by construction, since $K_{\rm Dir}(\varnothing)$ is a closed submanifold of it and the prescribed Dirichlet data are its induced metric. These data are CRT invariant, as must be the case for a norm. However, the saddle is not.
To see this, choose a point $p$ on the Lorentzian portion of $K_{\rm Dir}(\varnothing)$ with nonzero mean curvature $K(p)\neq 0$.
Because the saddle is Lorentzian in a neighborhood of $p$, the corresponding boundary extrinsic curvature is imaginary.
CRT invariance would therefore require
\begin{equation}\label{eq-KtominusK}
K(\boldsymbol{\Theta}(p)) = -K(p).
\end{equation}
But for the BTZ saddle under discussion, the two halves are related by the left--right reflection of the same future Cauchy slice, and hence instead satisfy
\begin{equation}\label{eq-KtoplusK}
K(\boldsymbol{\Theta}(p)) = K(p).
\end{equation}
Thus the saddle spontaneously breaks the boundary CRT symmetry. It follows that applying $\boldsymbol{\Theta}$ produces a second saddle obeying the same Dirichlet boundary conditions, whose action is the complex conjugate of the first. At leading order in $G_N$, this yields Eq.~\eqref{eq:general-CRT-pair} and can therefore give a negative norm.

\paragraph{Generic mixed-conformal boundary data from the Wall saddle}

We will now extract the mixed-conformal boundary data, rather than the Dirichlet boundary data, from the Wall saddle, on the same manifolds $J(\sigma)$ and $J^*(\bar\sigma)$ as before. 
Let $K_{\rm right}$ and $K_{\rm left}$ denote the traces of the extrinsic curvature on the Lorentzian piece of $J(\sigma)$ and of $J^*(\bar\sigma)$, respectively.

Left--right symmetry gives $K_{\rm left}(\boldsymbol{\Theta}(p)) = K_{\rm right}(p)$ for all $p$ on the Lorentzian portion of $J(\sigma)$. There $K_{\rm right}(p)$ is purely imaginary, so a norm would instead require $K_{\rm left}(\boldsymbol{\Theta}(p)) = \overline{K_{\rm right}(p)} = -K_{\rm right}(p)$. For a generic choice of Lorentzian slice, $K_{\rm right}(p)$ does not vanish. Hence the values of $K$ induced on $J(\sigma)$ and $J^*(\bar\sigma)$ are generically not related by CRT conjugation. It follows that the mixed-conformal data induced on $J(\sigma)$ and $J^*(\bar\sigma)$ are incompatible with a norm interpretation of the GPI. Instead, the GPI with the Wall saddles computes merely an inner product between two different states, and its evident negativity presents no problem.

\paragraph{Mixed-conformal norm data from the Wall saddle}
For the induced mixed-conformal data to be CRT invariant, and thus to have an interpretation as a norm, we must choose a non-generic Lorentzian slice with $K_{\rm right} \equiv 0$. We will now show that the GPI with the resulting mixed-conformal conditions admits a CRT-invariant solution which dominates over the Wall saddle for all choices of parameters, ensuring positivity.


We may decompose
\begin{equation}
J(\sigma)=J_0\cup_{\zeta_{0\kappa}}J_\kappa,
\qquad
\partial J(\sigma)=\sigma=\chi\sqcup\chi_c,
\end{equation}
into a Lorentzian arm $J_0$ with $K=0$, and a Euclidean constant-radius arm $J_\kappa$ with $K=\kappa$, joined at the internal joint $\zeta_{0\kappa}$. The boundary component $\chi$ is the Lorentzian midpoint on the $K=0$ arm, while $\chi_c$ is the Euclidean midpoint on the $K=\kappa$ arm. When equipped with mixed-conformal data we denote the manifold by $\mathcal J(\sigma)$.

We parameterize the Wall saddle by $r_h$, $R_c$, and $R_0$: $r_h$ is the horizon radius in the Wall filling, $R_c$ is the cutoff radius and the radius of $\chi_c$, and $R_0$ is the radius of $\chi$, with $r_h/\sqrt{2}<R_0<r_h<R_c$.
The prescribed trace is $K=0$ on $J_0$ and $K=\kappa$ on $J_\kappa$, where
\begin{equation}
\kappa
=
\frac{R_c}{\sqrt{R_c^2-r_h^2}}
+
\frac{\sqrt{R_c^2-r_h^2}}{R_c}.
\label{eq:wall-kappa-rewrite}
\end{equation}
The conformal classes of $J_0$ and $J_\kappa$ are fixed by their moduli $m_0$ and $m_\kappa$, respectively, in the convention $m=\Delta x/(2\pi)$.
For the data considered here, these moduli are
\begin{equation}
\begin{aligned}
m_\kappa
&=
\frac{\sqrt{R_c^2-r_h^2}}{4r_hR_c},
\\
m_0
&=
\frac{1}{2\pi}
\int_{R_0}^{R_c}
\frac{dr}{\sqrt{(r^2-R_0^2)(r^2-r_h^2+R_0^2)}}.
\end{aligned}
\label{eq:wall-target-moduli-rewrite}
\end{equation}
The Dirichlet part of the mixed-conformal data fixes
\begin{equation*}
\gamma_\chi=R_0^2d\phi^2,
\qquad
\gamma_{\chi_c}=\gamma_{\zeta_{0\kappa}}=R_c^2d\phi^2,
\qquad
\phi\sim\phi+2\pi.
\end{equation*}
Together, these conditions specify the mixed-conformal data on the preparation EDM.
To form the norm, we glue the original and involuted preparations across both boundary components:
\begin{equation*}
\mathcal J_1(\varnothing)
=
(\mathcal J^*)_A(\bar\sigma)\,
\mathcal J^A(\sigma).
\end{equation*}
 The Wall geometry and the fully Euclidean geometry constructed below are two saddles filling this same closed EDM; hence both must reproduce the same $m_0$, $m_\kappa$, traces, and joint metrics.
For the Wall filling, let $I_{\rm Wall}$ denote the action of its saddle, so that 
\begin{equation}
    \operatorname{Re}I_{\rm Wall}=-{\pi r_h\over 8G_N} -{\pi R_c \over 4G_N},
\end{equation}
where the first term comes from the Euclidean bulk and conformal-boundary terms and the second from the two cutoff joints.

\begin{figure}[tbp]
    \centering
    \includegraphics[width=0.704\linewidth]{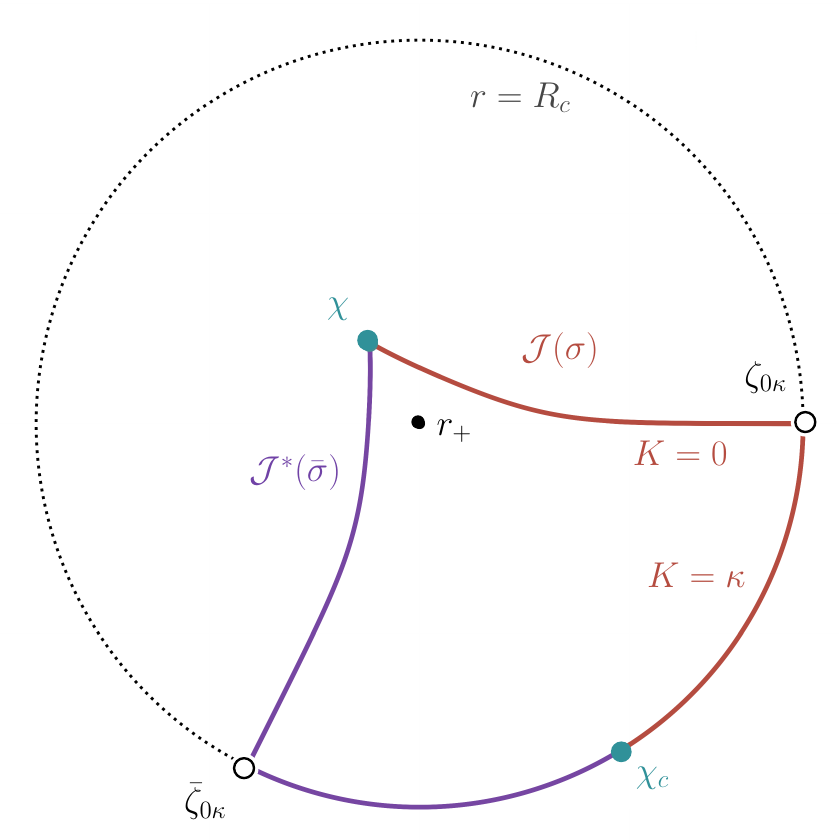}
    \begin{figurecaptionblock}
    \caption[Fully Euclidean CRT-invariant saddle]{The dominant Euclidean CRT-invariant saddle-point contribution to the norm.
    The mixed-conformal elliptic boundary data are identical to those that admit the CRT-breaking saddle shown in Fig.~\ref{fig:wall-norm-schematic}.
    The plotted example uses $(r_h,R_c,R_0)=(2,20,3/2)$. }
    \label{fig:wall-fully-euclidean-rewrite}
    \end{figurecaptionblock}
\end{figure}

We now consider the second filling, a CRT-invariant saddle in which all four arms lie in a Euclidean BTZ geometry.
Let $r_+$ be the horizon radius of this geometry, which need not equal the Wall radius $r_h$.
The four arms are grouped into the original and involuted preparations:
\begin{equation*}
J(\sigma)=J_0\cup_{\zeta_{0\kappa}}J_\kappa,
\qquad
J^*(\bar\sigma)=J_0^*\cup_{\bar\zeta_{0\kappa}}J_\kappa^*.
\end{equation*}
We glue these two preparations across $\chi$ and $\chi_c$, as shown in Fig.~\ref{fig:wall-fully-euclidean-rewrite}.
We first solve the embedding equation for each arm and then require the four arms to form one closed curve in the Euclidean BTZ plane.
For an arm with constant value $K=0$ or $K=\kappa$, define the integration constant $\mathcal E_K$ and the radial function $F_K(r)=
r^2(r^2-r_+^2)
-
\left(\frac{K}{2}r^2+\mathcal E_K\right)^2$ so that
\begin{equation}
\begin{aligned}
ds_{J_K}^2
&=
\frac{r^2\,dr^2}{F_K(r)}
+r^2d\phi^2,
\\
\frac{d\tau}{dr}
&=
\pm
\frac{\frac{K}{2}r^2+\mathcal E_K}
{(r^2-r_+^2)\sqrt{F_K(r)}}.
\end{aligned}
\label{eq:wall-FK-rewrite}
\end{equation}
The sign in the last line fixes the orientation of a radial branch, while the condition $F_K(r)\geq0$ determines the allowed radial interval.

To be more specific, we inspect the two $K=0$ arms $J_0$ and $J_0^*$.
Each arm runs between the circles at $R_0$ and $R_c$ and passes smoothly through a turning radius $r_{\mathrm t,0}<R_0$.
The two arms are related by reflection and therefore have the same $r_{\mathrm t,0}$ and $\mathcal E_0$.
The turning-point condition fixes
\begin{equation}
\begin{aligned}
0<r_+<r_{\mathrm t,0}<R_0,
\qquad
\mathcal E_0
&=
r_{\mathrm t,0}
\sqrt{r_{\mathrm t,0}^2-r_+^2},
\\
F_0(r)
&=
r^2(r^2-r_+^2)-\mathcal E_0^2.
\end{aligned}
\label{eq:wall-K0-turn-rewrite}
\end{equation}
Each $K=0$ arm contains one radial branch from $r_{\mathrm t,0}$ to $R_c$ and one from $r_{\mathrm t,0}$ to $R_0$.
The modulus of either arm and the total Euclidean time span of the two arms are
\begin{equation}
\begin{aligned}
m_0^{\rm Euc}
&=
\frac{1}{2\pi}
\left[
\int_{r_{\mathrm t,0}}^{R_c}
\frac{d\rho}{\sqrt{F_0(\rho)}}
+
\int_{r_{\mathrm t,0}}^{R_0}
\frac{d\rho}{\sqrt{F_0(\rho)}}
\right],
\\
\Delta\tau_0
&=
2
\left[
\int_{r_{\mathrm t,0}}^{R_c}
\frac{\mathcal E_0\,d\rho}{(\rho^2-r_+^2)\sqrt{F_0(\rho)}}
+
\int_{r_{\mathrm t,0}}^{R_0}
\frac{\mathcal E_0\,d\rho}{(\rho^2-r_+^2)\sqrt{F_0(\rho)}}
\right].
\end{aligned}
\label{eq:wall-K0-mod-time-rewrite}
\end{equation}

We next inspect the embedding of the two $K=\kappa$ arms $J_\kappa$ and $J_\kappa^*$.
On either arm,
\begin{equation}
\begin{aligned}
F_\kappa(r)
&=
r^2(r^2-r_+^2)
-
\left(\frac{\kappa}{2}r^2+\mathcal E_\kappa\right)^2,
\\
F_\kappa(r_{\mathrm t,\kappa})
&=0,
\qquad
R_c<r_{\mathrm t,\kappa}.
\end{aligned}
\label{eq:wall-Fkappa-rewrite}
\end{equation}
Each $K=\kappa$ arm runs from $R_c$ to $r_{\mathrm t,\kappa}$ and back to $R_c$.
The modulus of either arm and the total Euclidean time span of the two arms are
\begin{equation}
\begin{aligned}
m_\kappa^{\rm Euc}
&=
\frac{1}{\pi}
\int_{R_c}^{r_{\mathrm t,\kappa}}
\frac{dr}{\sqrt{F_\kappa(r)}},
\\
\Delta\tau_\kappa
&=
4
\int_{R_c}^{r_{\mathrm t,\kappa}}
\frac{\frac{\kappa}{2}r^2+\mathcal E_\kappa}
{(r^2-r_+^2)\sqrt{F_\kappa(r)}}
\,dr.
\end{aligned}
\label{eq:wall-kappa-mod-time-rewrite}
\end{equation}

At this point, each arm solves its embedding equation.
The four arms form a saddle only if they reproduce $m_0$ and $m_\kappa$ and close the thermal circle.
These requirements give the three simultaneous equations
\begin{align}
m_0^{\rm Euc}(r_+,r_{\mathrm t,0})
&=m_0,
\nonumber\\
m_\kappa^{\rm Euc}(r_+,\mathcal E_\kappa)
&=m_\kappa,
\nonumber\\
\Delta\tau_0(r_+,r_{\mathrm t,0})
+
\Delta\tau_\kappa(r_+,\mathcal E_\kappa)
&=\frac{2\pi}{r_+}.
\label{eq:wall-matching-rewrite}
\end{align}
The first two lines match the two conformal moduli, while the last closes the thermal circle.
They must be solved simultaneously for $r_+$, $r_{\mathrm t,0}$, and $\mathcal E_\kappa$; the value of $\mathcal E_0$ then follows from Eq.~\eqref{eq:wall-K0-turn-rewrite}.
For the Euclidean solution,
\begin{equation*}
0<r_+<r_{\mathrm t,0}<R_0<r_h<R_c<r_{\mathrm t,\kappa},
\qquad
\kappa>2,
\qquad
\mathcal E_0>0>\mathcal E_\kappa.
\end{equation*}
The resulting closed embedding is shown in Fig.~\ref{fig:wall-fully-euclidean-rewrite}.

We now compare the actions of the Wall and fully Euclidean saddles.
This comparison must include the Hayward contributions from the joints shown in Fig.~\ref{fig:wall-fully-euclidean-rewrite}.
With exterior angle $\Theta_\zeta=\pi-\alpha_\zeta$ at a circular joint $\zeta$, our Hayward convention is
\begin{equation}
G_N I_H(\zeta)=-\frac{r_\zeta\Theta_\zeta}{4}.
\label{eq:wall-hayward-joint-rewrite}
\end{equation}
At each internal joint $\zeta_{0\kappa}$ and $\bar\zeta_{0\kappa}$, a $K=0$ arm meets a $K=\kappa$ arm at $R_c$.
At $\chi_c$, the two $K=\kappa$ arms form a concave joint at $R_c$; at $\chi$, the two $K=0$ arms form a convex joint at $R_0$.
Their exterior angles are
\begin{equation}
\begin{aligned}
\Theta_{0\kappa}
&=
\frac{\pi}{2}
-
\arctan\frac{\mathcal E_0}{\sqrt{F_0(R_c)}}
-
\arctan\frac{\sqrt{F_\kappa(R_c)}}{\frac{\kappa}{2}R_c^2+\mathcal E_\kappa},
\\
\Theta_{\kappa\kappa}
&=
-2\arctan\frac{\sqrt{F_\kappa(R_c)}}{\frac{\kappa}{2}R_c^2+\mathcal E_\kappa},
\\
\Theta_{00}
&=
2\arctan\frac{\sqrt{F_0(R_0)}}{\mathcal E_0}.
\end{aligned}
\label{eq:wall-hayward-angles-rewrite}
\end{equation}
Here $\Theta_{00}$ is the angle at $\chi$ and $\Theta_{\kappa\kappa}$ is the angle at $\chi_c$; their subscripts record the values of $K$ on the two arms that meet there.
The negative sign of $\Theta_{\kappa\kappa}$ records that the joint between the two $K=\kappa$ arms is concave relative to the Euclidean region included in the saddle.

\begin{figure}
    \centering
    \includegraphics[width=0.78\linewidth]{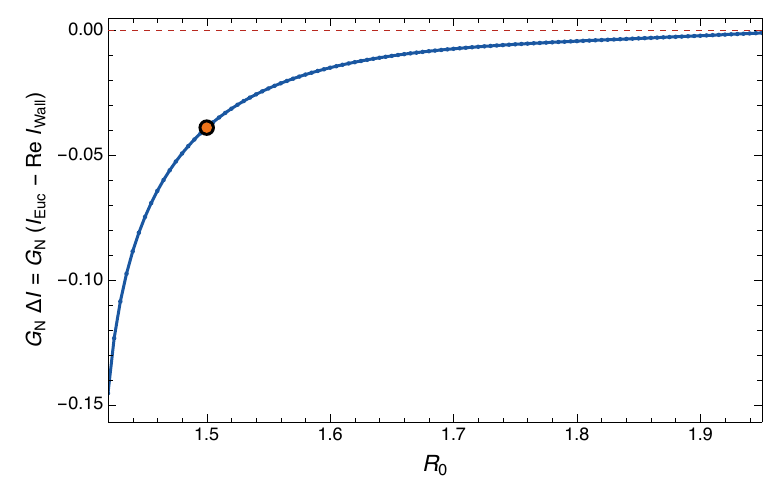}
    \begin{figurecaptionblock}
    \caption[Action difference along the parameter scan]{The action difference $\Delta I=I_{\rm Euc}-\operatorname{Re}I_{\rm Wall}$ along a parameter scan with $R_c=20$, $r_h=2$, and $r_h/\sqrt{2}<R_0<r_h$.
    At each displayed value of $R_0$, Eq.~\eqref{eq:wall-matching-rewrite} is solved again on the chosen outward branch before the smooth and Hayward contributions are evaluated.
    The orange point marks the reference solution at $R_0=3/2$, where $G_N\Delta I\simeq-0.0389053$.
    The red dashed line at $\Delta I=0$ indicates equality of the real parts of the two actions; since $G_N\Delta I<0$ along the curve, the Euclidean saddle always dominates.
    }
    \label{fig:wall-action-difference-rewrite}
    \end{figurecaptionblock}
\end{figure}

The Einstein--Hilbert and conformal-boundary terms define the smooth contribution
\begin{equation}
G_N I_{\rm smooth}^{\rm Euc}
=
\pi
\left(
\mathcal E_0m_0^{\rm Euc}
+
\mathcal E_\kappa m_\kappa^{\rm Euc}
\right),
\label{eq:wall-smooth-action-rewrite}
\end{equation}
and the sum of the Hayward contributions is
\begin{equation}
G_N I_H^{\rm Euc}
=
-\frac{1}{4}
\left(
2R_c\Theta_{0\kappa}
+
R_c\Theta_{\kappa\kappa}
+
R_0\Theta_{00}
\right).
\label{eq:wall-hayward-action-rewrite}
\end{equation}
We define $I_{\rm Euc}=I_{\rm smooth}^{\rm Euc}+I_H^{\rm Euc}$.
For the reference example, the common cutoff-dependent term cancels against the same term in $\operatorname{Re}I_{\rm Wall}$, leaving
\begin{equation}
G_N\left(I_{\rm Euc}-\operatorname{Re}I_{\rm Wall}\right)<0.
\label{eq:wall-action-difference-rewrite}
\end{equation}
The CRT-invariant saddle therefore has smaller real action than either member of the CRT-breaking pair for this boundary condition.\footnote{\interlinepenalty=10000\relax The matching equations also admit a different ``long branch'' with $r_{\mathrm t,\kappa}\gg R_c$ whose turning radius and action do not have the well-behaved large-$R_c$ behavior of the branch used here. For example, as $R_c\to\infty$, the maximal area grows as $2\pi r_{\mathrm t,\kappa}\sim16R_c^2\log(R_c/r_h)/r_h$, and the leading action difference from $\operatorname{Re}I_{\rm Wall}$ is $-2r_h\log^2(R_c/r_h)/(\pi G_N)$.
We therefore assume the long branch saddle does not contribute, though the positivity conclusion does not rely on this assumption since we already have a ``saving'' geometry described in the text.}
Repeating the numerical solution while varying $R_0$, we find the same inequality at every value for which the matching equations admit a solution, as shown in Fig.~\ref{fig:wall-action-difference-rewrite}.\footnote{An alternative solution satisfying the equal-span condition $\Delta\tau_0=\Delta\tau_\kappa$ exists in part of parameter space, but its joint at $R_0$ is concave relative to the Euclidean region included in the saddle.
Its Hayward contribution raises the action, so it is subleading to the saddle used in the main comparison.}

This saddle removes the apparent positivity violation in the configurations examined here: although the combined contribution of the CRT-breaking pair can be negative, it is subleading to the positive contribution from the CRT-invariant saddle.
The comparison establishes this result for this specific example over the scanned parameter range, rather than providing a general proof of positivity for all states based on mixed-conformal data.\footnote{Within the Dirichlet family considered here, the full induced metric fixes both the radial geometry and its domain $R_0\leq r\leq R_c$.
Appending a branch below $R_0$ would change the Dirichlet data, and neither CRT-invariant candidate satisfies the same boundary data within this family.
This does not exclude other CRT-invariant saddles for more general Dirichlet data.}

\paragraph{Acknowledgements} We thank Mina Aganagic, Luca Iliesiu, Henry Maxfield, and Pratik Rath for discussions.
This work was supported in part by the Leinweber Institutes for Theoretical Physics at UC Berkeley and Stanford; and by the Department of Energy, Office of Science, Office of High Energy Physics through award DE-SC0025293 and QuantISED award DE-SC0019380.
ASM is supported by the Department of Energy through awards DE-SC0019380 and DE-FOA0002563, by AFOSR award FA9550-22-1-0098, and by a Sloan Fellowship.

\bibliographystyle{JHEP}
\bibliography{Refs}

@article{Banihashemi:2024flat,
    author = {Banihashemi, Batoul and Shaghoulian, Edgar and Shashi, Sanjit},
    title = {Flat space gravity at finite cutoff},
    eprint = {2409.07643},
    archivePrefix = {arXiv},
    primaryClass = {hep-th},
    doi = {10.1088/1361-6382/ada2d7},
    journal = {Class. Quant. Grav.},
    volume = {42},
    number = {3},
    pages = {035010},
    year = {2025}
}

@article{Banihashemi:2025thermal,
    author = {Banihashemi, Batoul and Shaghoulian, Edgar and Shashi, Sanjit},
    title = {Thermal effective actions from conformal boundary conditions in gravity},
    eprint = {2503.17471},
    archivePrefix = {arXiv},
    primaryClass = {hep-th},
    doi = {10.1088/1361-6382/adee72},
    journal = {Class. Quant. Grav.},
    volume = {42},
    number = {15},
    pages = {155004},
    year = {2025}
}

@article{Allameh:2025timelike,
    author = {Allameh, Kuroush and Shaghoulian, Edgar},
    title = {Timelike {Liouville} theory and {AdS}$_3$ gravity at finite cutoff},
    eprint = {2508.03236},
    archivePrefix = {arXiv},
    primaryClass = {hep-th},
    year = {2025}
}

@article{Gibbons:1976ue,
    author = "Gibbons, G. W. and Hawking, S. W.",
    title = "{Action Integrals and Partition Functions in Quantum Gravity}",
    doi = "10.1103/PhysRevD.15.2752",
    journal = "Phys. Rev. D",
    volume = "15",
    pages = "2752--2756",
    year = "1977"
}

@article{Hawking:1982dh,
    author = "Hawking, S. W. and Page, Don N.",
    title = "{Thermodynamics of Black Holes in anti-De Sitter Space}",
    doi = "10.1007/BF01208266",
    journal = "Commun. Math. Phys.",
    volume = "87",
    pages = "577--588",
    year = "1983"
}

@article{Gubser:1998bc,
    author = "Gubser, S. S. and Klebanov, Igor R. and Polyakov, Alexander M.",
    title = "{Gauge theory correlators from noncritical string theory}",
    eprint = "hep-th/9802109",
    archivePrefix = "arXiv",
    reportNumber = "PUPT-1767",
    doi = "10.1016/S0370-2693(98)00377-3",
    journal = "Phys. Lett. B",
    volume = "428",
    pages = "105--114",
    year = "1998"
}

@article{Witten:1998qj,
    author = "Witten, Edward",
    title = "{Anti-de Sitter space and holography}",
    eprint = "hep-th/9802150",
    archivePrefix = "arXiv",
    reportNumber = "IASSNS-HEP-98-15",
    doi = "10.4310/ATMP.1998.v2.n2.a2",
    journal = "Adv. Theor. Math. Phys.",
    volume = "2",
    pages = "253--291",
    year = "1998"
}

@article{Bousso:2023sya,
    author = "Bousso, Raphael and Penington, Geoff",
    title = "{Holograms in our world}",
    eprint = "2302.07892",
    archivePrefix = "arXiv",
    primaryClass = "hep-th",
    doi = "10.1103/PhysRevD.108.046007",
    journal = "Phys. Rev. D",
    volume = "108",
    number = "4",
    pages = "046007",
    year = "2023"
}

@article{Almheiri:2014lwa,
    author = "Almheiri, Ahmed and Dong, Xi and Harlow, Daniel",
    title = "{Bulk Locality and Quantum Error Correction in AdS/CFT}",
    eprint = "1411.7041",
    archivePrefix = "arXiv",
    primaryClass = "hep-th",
    reportNumber = "SU-ITP-14-30",
    doi = "10.1007/JHEP04(2015)163",
    journal = "JHEP",
    volume = "04",
    pages = "163",
    year = "2015"
}

@article{Bousso:2022hlz,
    author = "Bousso, Raphael and Penington, Geoff",
    title = "{Entanglement Wedges for Gravitating Regions}",
    eprint = "2208.04993",
    archivePrefix = "arXiv",
    primaryClass = "hep-th",
    month = "8",
    year = "2022"
}

@article{Gorbenko:2018oov,
    author = "Gorbenko, Victor and Silverstein, Eva and Torroba, Gonzalo",
    title = "{dS/dS and $ T\overline{T} $}",
    eprint = "1811.07965",
    archivePrefix = "arXiv",
    primaryClass = "hep-th",
    doi = "10.1007/JHEP03(2019)085",
    journal = "JHEP",
    volume = "03",
    pages = "085",
    year = "2019"
}

@article{Lewkowycz:2019xse,
    author = "Lewkowycz, Aitor and Liu, Junyu and Silverstein, Eva and Torroba, Gonzalo",
    title = "{$ T\overline{T} $ and EE, with implications for (A)dS subregion encodings}",
    eprint = "1909.13808",
    archivePrefix = "arXiv",
    primaryClass = "hep-th",
    reportNumber = "CALT-TH-2019--031",
    doi = "10.1007/JHEP04(2020)152",
    journal = "JHEP",
    volume = "04",
    pages = "152",
    year = "2020"
}

@article{Hartman:2018tkw,
    author = "Hartman, Thomas and Kruthoff, Jorrit and Shaghoulian, Edgar and Tajdini, Amirhossein",
    title = "{Holography at finite cutoff with a $T^2$ deformation}",
    eprint = "1807.11401",
    archivePrefix = "arXiv",
    primaryClass = "hep-th",
    doi = "10.1007/JHEP03(2019)004",
    journal = "JHEP",
    volume = "03",
    pages = "004",
    year = "2019"
}

@article{McGough:2016lol,
    author = "McGough, Lauren and Mezei, M\'ark and Verlinde, Herman",
    title = "{Moving the CFT into the bulk with $ T\overline{T} $}",
    eprint = "1611.03470",
    archivePrefix = "arXiv",
    primaryClass = "hep-th",
    doi = "10.1007/JHEP04(2018)010",
    journal = "JHEP",
    volume = "04",
    pages = "010",
    year = "2018"
}

@article{Almheiri:2012rt,
    author = "Almheiri, Ahmed and Marolf, Donald and Polchinski, Joseph and Sully, James",
    title = "{Black Holes: Complementarity or Firewalls?}",
    eprint = "1207.3123",
    archivePrefix = "arXiv",
    primaryClass = "hep-th",
    doi = "10.1007/JHEP02(2013)062",
    journal = "JHEP",
    volume = "02",
    pages = "062",
    year = "2013"
}

@article{Pastawski:2015qua,
    author = "Pastawski, Fernando and Yoshida, Beni and Harlow, Daniel and Preskill, John",
    title = "{Holographic quantum error-correcting codes: Toy models for the bulk/boundary correspondence}",
    eprint = "1503.06237",
    archivePrefix = "arXiv",
    primaryClass = "hep-th",
    doi = "10.1007/JHEP06(2015)149",
    journal = "JHEP",
    volume = "06",
    pages = "149",
    year = "2015"
}

@article{Swingle:2009bg,
    author = "Swingle, Brian",
    title = "{Entanglement Renormalization and Holography}",
    eprint = "0905.1317",
    archivePrefix = "arXiv",
    primaryClass = "cond-mat.str-el",
    doi = "10.1103/PhysRevD.86.065007",
    journal = "Phys. Rev. D",
    volume = "86",
    pages = "065007",
    year = "2012"
}

@article{Lewkowycz:2013nqa,
    author = "Lewkowycz, Aitor and Maldacena, Juan",
    title = "{Generalized gravitational entropy}",
    eprint = "1304.4926",
    archivePrefix = "arXiv",
    primaryClass = "hep-th",
    doi = "10.1007/JHEP08(2013)090",
    journal = "JHEP",
    volume = "08",
    pages = "090",
    year = "2013"
}

@article{Dong:2016fnf,
    author = "Dong, Xi",
    title = "{The Gravity Dual of R\'enyi Entropy}",
    eprint = "1601.06788",
    archivePrefix = "arXiv",
    primaryClass = "hep-th",
    reportNumber = "SU-ITP-16/01",
    doi = "10.1038/ncomms12472",
    journal = "Nature Commun.",
    volume = "7",
    pages = "12472",
    year = "2016"
}

@article{Ryu:2006bv,
    author = "Ryu, Shinsei and Takayanagi, Tadashi",
    title = "{Holographic derivation of entanglement entropy from AdS/CFT}",
    eprint = "hep-th/0603001",
    archivePrefix = "arXiv",
    reportNumber = "NSF-KITP-06-11",
    doi = "10.1103/PhysRevLett.96.181602",
    journal = "Phys. Rev. Lett.",
    volume = "96",
    pages = "181602",
    year = "2006"
}

@article{Hubeny:2007xt,
    author = "Hubeny, Veronika E. and Rangamani, Mukund and Takayanagi, Tadashi",
    title = "{A Covariant holographic entanglement entropy proposal}",
    eprint = "0705.0016",
    archivePrefix = "arXiv",
    primaryClass = "hep-th",
    reportNumber = "DCPT-07-13, KUNS-2069",
    doi = "10.1088/1126-6708/2007/07/062",
    journal = "JHEP",
    volume = "07",
    pages = "062",
    year = "2007"
}

@article{Akers:2019wxj,
    author = "Akers, Chris and Leichenauer, Stefan and Levine, Adam",
    title = "{Large Breakdowns of Entanglement Wedge Reconstruction}",
    eprint = "1908.03975",
    archivePrefix = "arXiv",
    primaryClass = "hep-th",
    doi = "10.1103/PhysRevD.100.126006",
    journal = "Phys. Rev. D",
    volume = "100",
    number = "12",
    pages = "126006",
    year = "2019"
}

@article{Hayden:2016cfa,
    author = "Hayden, Patrick and Nezami, Sepehr and Qi, Xiao-Liang and Thomas, Nathaniel and Walter, Michael and Yang, Zhao",
    title = "{Holographic duality from random tensor networks}",
    eprint = "1601.01694",
    archivePrefix = "arXiv",
    primaryClass = "hep-th",
    doi = "10.1007/JHEP11(2016)009",
    journal = "JHEP",
    volume = "11",
    pages = "009",
    year = "2016"
}

@article{Wall:2012uf,
    author = "Wall, Aron C.",
    title = "{Maximin Surfaces, and the Strong Subadditivity of the Covariant Holographic Entanglement Entropy}",
    eprint = "1211.3494",
    archivePrefix = "arXiv",
    primaryClass = "hep-th",
    doi = "10.1088/0264-9381/31/22/225007",
    journal = "Class. Quant. Grav.",
    volume = "31",
    number = "22",
    pages = "225007",
    year = "2014"
}

@article{Marolf:2020xie,
    author = "Marolf, Donald and Maxfield, Henry",
    title = "{Transcending the ensemble: baby universes, spacetime wormholes, and the order and disorder of black hole information}",
    eprint = "2002.08950",
    archivePrefix = "arXiv",
    primaryClass = "hep-th",
    doi = "10.1007/JHEP08(2020)044",
    journal = "JHEP",
    volume = "08",
    pages = "044",
    year = "2020"
}

@article{Saad:2019lba,
    author = "Saad, Phil and Shenker, Stephen H. and Stanford, Douglas",
    title = "{JT gravity as a matrix integral}",
    eprint = "1903.11115",
    archivePrefix = "arXiv",
    primaryClass = "hep-th",
    year = "2019"
}

@article{Atiyah:1988,
    author = "Atiyah, Michael",
    title = "{Topological quantum field theories}",
    doi = "10.1007/BF02698547",
    journal = "Inst. Hautes Etudes Sci. Publ. Math.",
    volume = "68",
    pages = "175--186",
    year = "1988"
}

@incollection{Segal:2004,
    author = "Segal, Graeme",
    title = "{The definition of conformal field theory}",
    booktitle = "{Topology, Geometry and Quantum Field Theory}",
    editor = "Tillmann, Ulrike",
    doi = "10.1017/CBO9780511526398.020",
    series = "London Mathematical Society Lecture Note Series",
    volume = "308",
    pages = "421--577",
    publisher = "Cambridge University Press",
    year = "2004"
}

@article{Anderson:2006bvp,
    author = "Anderson, Michael T.",
    title = "{On boundary value problems for Einstein metrics}",
    eprint = "math/0612647",
    archivePrefix = "arXiv",
    primaryClass = "math.DG",
    doi = "10.2140/gt.2008.12.2009",
    journal = "Geom. Topol.",
    volume = "12",
    pages = "2009--2045",
    year = "2008"
}

@article{Witten:2018lgb,
    author = "Witten, Edward",
    title = "{A note on boundary conditions in Euclidean gravity}",
    eprint = "1805.11559",
    archivePrefix = "arXiv",
    primaryClass = "hep-th",
    doi = "10.1142/S0129055X21400043",
    journal = "Rev. Math. Phys.",
    volume = "33",
    number = "10",
    pages = "2140004",
    year = "2021"
}

@article{Hayward:1993my,
    author = "Hayward, G.",
    title = "{Gravitational action for space-times with nonsmooth boundaries}",
    doi = "10.1103/PhysRevD.47.3275",
    journal = "Phys. Rev. D",
    volume = "47",
    pages = "3275--3280",
    year = "1993"
}

@article{Araujo-Regado:2022gvw,
    author = "Araujo-Regado, Goncalo and Khan, Rifath and Wall, Aron C.",
    title = "{Cauchy slice holography: a new AdS/CFT dictionary}",
    eprint = "2204.00591",
    archivePrefix = "arXiv",
    primaryClass = "hep-th",
    doi = "10.1007/JHEP03(2023)026",
    journal = "JHEP",
    volume = "03",
    number = "2023",
    pages = "026",
    year = "2023"
}

@article{MarolfColafranceschi,
    author = "Colafranceschi, Eugenia and Dong, Xi and Marolf, Donald and Wang, Zhencheng",
    title = "{Algebras and Hilbert spaces from gravitational path integrals. Understanding Ryu-Takayanagi/HRT as entropy without AdS/CFT}",
    eprint = "2310.02189",
    archivePrefix = "arXiv",
    primaryClass = "hep-th",
    doi = "10.1007/JHEP10(2024)063",
    journal = "JHEP",
    volume = "10",
    pages = "063",
    year = "2024"
}

@article{Wall:2021bxi,
    author = "Wall, Aron C.",
    title = "{Violation of unitarity in gravitational subregions}",
    eprint = "2104.03253",
    archivePrefix = "arXiv",
    primaryClass = "gr-qc",
    doi = "10.1142/S0218271821420141",
    journal = "Int. J. Mod. Phys. D",
    volume = "30",
    number = "14",
    pages = "2142014",
    year = "2021"
}

@book{PoissonBook,
    author = "Poisson, Eric",
    title = "{A Relativist's Toolkit: The Mathematics of Black-Hole Mechanics}",
    publisher = "Cambridge University Press",
    doi = "10.1017/CBO9780511606601",
    year = "2004"
}

@article{Abdalla:2026nbp,
    author = "Abdalla, Ahmed I. and Antonini, Stefano and Bousso, Raphael and Iliesiu, Luca V. and Levine, Adam and Shahbazi-Moghaddam, Arvin",
    title = "{Consistent Evaluation of the No-Boundary Proposal}",
    eprint = "2602.02682",
    archivePrefix = "arXiv",
    primaryClass = "hep-th",
    year = "2026"
}

@article{Usatyuk:2024mzs,
    author = "Usatyuk, Mykhaylo and Wang, Zi-Yue and Zhao, Ying",
    title = "{Closed universes in two dimensional gravity}",
    eprint = "2402.00098",
    archivePrefix = "arXiv",
    primaryClass = "hep-th",
    doi = "10.21468/SciPostPhys.17.2.051",
    journal = "SciPost Phys.",
    volume = "17",
    number = "2",
    pages = "051",
    year = "2024"
}

@article{Usatyuk:2024isz,
    author = "Usatyuk, Mykhaylo and Zhao, Ying",
    title = "{Closed universes, factorization, and ensemble averaging}",
    eprint = "2403.13047",
    archivePrefix = "arXiv",
    primaryClass = "hep-th",
    doi = "10.1007/JHEP02(2025)052",
    journal = "JHEP",
    volume = "02",
    pages = "052",
    year = "2025"
}

@article{Harlow:2025pvj,
    author = "Harlow, Daniel and Usatyuk, Mykhaylo and Zhao, Ying",
    title = "{Quantum mechanics and observers for gravity in a closed universe}",
    eprint = "2501.02359",
    archivePrefix = "arXiv",
    primaryClass = "hep-th",
    reportNumber = "MIT-CTP/5824",
    doi = "10.1007/JHEP02(2026)108",
    journal = "JHEP",
    volume = "02",
    pages = "108",
    year = "2026"
}

@article{Abdalla:2025gzn,
    author = "Abdalla, Ahmed I. and Antonini, Stefano and Iliesiu, Luca V. and Levine, Adam",
    title = "{The gravitational path integral from an observer{\textquoteright}s point of view}",
    eprint = "2501.02632",
    archivePrefix = "arXiv",
    primaryClass = "hep-th",
    doi = "10.1007/JHEP05(2025)059",
    journal = "JHEP",
    volume = "05",
    pages = "059",
    year = "2025"
}

@article{Harlow:2026hky,
    author = "Harlow, Daniel",
    title = "{Observers, {$\alpha$}-parameters, and the Hartle-Hawking state}",
    eprint = "2602.03835",
    archivePrefix = "arXiv",
    primaryClass = "hep-th",
    reportNumber = "MIT-CTP/5900",
    month = "2",
    year = "2026"
}

@article{Zhao:2026mpl,
    author = "Zhao, Ying",
    title = "{''It from Bit'': The Hartle-Hawking state and quantum mechanics for de Sitter observers}",
    eprint = "2602.05939",
    archivePrefix = "arXiv",
    primaryClass = "hep-th",
    reportNumber = "MIT-CTP/6000",
    month = "2",
    year = "2026"
}

@article{Nomura:2026igt,
    author = "Nomura, Yasunori and Ugajin, Tomonori",
    title = "{Physical Predictions in Closed Quantum Gravity}",
    eprint = "2602.13387",
    archivePrefix = "arXiv",
    primaryClass = "hep-th",
    reportNumber = "RIKEN-iTHEMS-Report-26",
    month = "2",
    year = "2026"
}

@article{shortpaper,
    author = "Bousso, Raphael and Kaya, Sami and Lin, Guanda and Shahbazi-Moghaddam, Arvin",
    title = "{Quantum State of a Gravitating Region}",
    eprint = "2605.28958",
    archivePrefix = "arXiv",
    primaryClass = "hep-th",
    year = "2026"
}

@article{Hartle:1983ai,
    author = "Hartle, J. B. and Hawking, S. W.",
    title = "{Wave Function of the Universe}",
    doi = "10.1103/PhysRevD.28.2960",
    journal = "Phys. Rev. D",
    volume = "28",
    pages = "2960--2975",
    year = "1983"
}

@article{Vidal:2005zs,
    author = "Vidal, Guifre",
    title = "{Entanglement Renormalization}",
    eprint = "cond-mat/0512165",
    archivePrefix = "arXiv",
    doi = "10.1103/PhysRevLett.99.220405",
    journal = "Phys. Rev. Lett.",
    volume = "99",
    pages = "220405",
    year = "2007"
}

@article{Vidal:2006qt,
    author = "Vidal, Guifre",
    title = "{Class of Quantum Many-Body States That Can Be Efficiently Simulated}",
    eprint = "quant-ph/0610099",
    archivePrefix = "arXiv",
    doi = "10.1103/PhysRevLett.101.110501",
    journal = "Phys. Rev. Lett.",
    volume = "101",
    pages = "110501",
    year = "2008"
}

@article{Caputa:2020fbc,
    author = "Caputa, Pawel and Kruthoff, Jorrit and Parrikar, Onkar",
    title = "{Building Tensor Networks for Holographic States}",
    eprint = "2012.05247",
    archivePrefix = "arXiv",
    primaryClass = "hep-th",
    doi = "10.1007/JHEP05(2021)009",
    journal = "JHEP",
    volume = "05",
    pages = "009",
    year = "2021",
    note = "[Erratum: JHEP 09, 112 (2022)]"
}

@article{Belin:2020oib,
    author = "Belin, Alexandre and Lewkowycz, Aitor and Sarosi, Gabor",
    title = "{Gravitational path integral from the $T^2$ deformation}",
    eprint = "2006.01835",
    archivePrefix = "arXiv",
    primaryClass = "hep-th",
    reportNumber = "CERN-TH-2020-085",
    doi = "10.1007/JHEP09(2020)156",
    journal = "JHEP",
    volume = "09",
    pages = "156",
    year = "2020"
}

@article{Chen:2025fwp,
    author = "Chen, Hong Zhe",
    title = "{Observers seeing gravitational Hilbert spaces: abstract sources for an abstract path integral}",
    eprint = "2505.15892",
    archivePrefix = "arXiv",
    primaryClass = "hep-th",
    doi = "10.1007/JHEP10(2025)139",
    journal = "JHEP",
    volume = "10",
    number = "2025",
    pages = "139",
    year = "2025"
}

@article{DysonKlebanSusskind,
    author = "Dyson, Lisa and Kleban, Matthew and Susskind, Leonard",
    title = "{Disturbing implications of a cosmological constant}",
    eprint = "hep-th/0208013",
    archivePrefix = "arXiv",
    reportNumber = "SU-ITP-02-25, MIT-CTP-3295",
    doi = "10.1088/1126-6708/2002/10/011",
    journal = "JHEP",
    volume = "10",
    pages = "011",
    year = "2002"
}

@article{York:1972sj,
  author  = {York, James W.},
  title   = {{Role of Conformal Three-Geometry in the Dynamics of Gravitation}},
  journal = {Phys. Rev. Lett.},
  volume  = {28},
  pages   = {1082--1085},
  year    = {1972},
  doi     = {10.1103/PhysRevLett.28.1082}
}

@article{Kaya:2025vof,
    author = "Kaya, Sami and Rath, Pratik and Ritchie, Kyle",
    title = "{Hollow-grams: generalized entanglement wedges from the gravitational path integral}",
    eprint = "2506.10064",
    archivePrefix = "arXiv",
    primaryClass = "hep-th",
    doi = "10.1007/JHEP09(2025)032",
    journal = "JHEP",
    volume = "09",
    number = "2025",
    pages = "032",
    year = "2025"
}

@misc{Balasubramanian:2023dpj,
    author = "Balasubramanian, Vijay and Cummings, Charlie",
    title = "{The entropy of finite gravitating regions}",
    eprint = "2312.08434",
    archivePrefix = "arXiv",
    primaryClass = "hep-th",
    month = "12",
    year = "2023"
}

@article{Candelas:1977rindler,
    author = "Candelas, P. and Deutsch, D.",
    title = "{On the vacuum stress induced by uniform acceleration or supporting the ether}",
    doi = "10.1098/rspa.1977.0057",
    journal = "Proc. Roy. Soc. Lond. A",
    volume = "354",
    pages = "79--99",
    year = "1977"
}

@article{Fulling:1973,
    author = "Fulling, Stephen A.",
    title = "{Nonuniqueness of Canonical Field Quantization in Riemannian Space-Time}",
    doi = "10.1103/PhysRevD.7.2850",
    journal = "Phys. Rev. D",
    volume = "7",
    pages = "2850--2862",
    year = "1973"
}

@article{Banados:1992wn,
  author        = {Banados, Maximo and Teitelboim, Claudio and Zanelli, Jorge},
  title         = {{The Black Hole in Three-Dimensional Space-Time}},
  journal       = {Phys. Rev. Lett.},
  volume        = {69},
  pages         = {1849--1851},
  year          = {1992},
  eprint        = {hep-th/9204099},
  archivePrefix = {arXiv},
  doi           = {10.1103/PhysRevLett.69.1849}
}

@article{Coleman:2020jte,
    author = "Coleman, Evan and Shyam, Vasudev",
    title = "{Conformal boundary conditions from cutoff AdS$_{3}$}",
    eprint = "2010.08504",
    archivePrefix = "arXiv",
    primaryClass = "hep-th",
    doi = "10.1007/JHEP09(2021)079",
    journal = "JHEP",
    volume = "09",
    pages = "079",
    year = "2021"
}

\end{document}